\documentclass[%
 reprint,
 superscriptaddress,
 amsmath,amssymb,
 aps,
 prl,
 longbibliography,
]{revtex4-2}

\usepackage{graphicx}
\usepackage{dcolumn}
\usepackage{bm}
\usepackage{url}
\usepackage[dvipsnames]{xcolor}
\usepackage{makecell}
\usepackage{physics}
\usepackage{upgreek}
\usepackage{float}
\usepackage{amsmath}
\usepackage{siunitx}
\usepackage{bbding}
\usepackage{multirow}
\usepackage{mathrsfs}
\usepackage{lineno}
\usepackage{hyperref}
\hypersetup{hidelinks}
\usepackage{bibunits}

\begin{document}

\title{Domain-Wall Mediated Polarization Switching in Ferroelectric AlScN: \\ Strain Relief and Field-Dependent Dynamics}

\author{Xiangyu Zheng}
\affiliation{Key Laboratory of Material Simulation Methods and Software of Ministry of Education, College of Physics, Jilin University, Changchun 130012, China}

\author{Charles Paillard}
\email{paillard@uark.edu}
\affiliation{Smart Ferroic Materials Center, Physics Department and Institute for Nanoscience and Engineering, University of Arkansas, Fayetteville, Arkansas 72701, USA}

\author{Dawei Wang}
\affiliation{Smart Ferroic Materials Center, Physics Department and Institute for Nanoscience and Engineering, University of Arkansas, Fayetteville, Arkansas 72701, USA}

\author{Peng Chen}
\affiliation{Guangdong Technion – Israel Institute of Technology, Guangdong 515063, China}
\affiliation{Department of Physics, Technion – Israel Institute of Technology, 32000 Haifa, Israel}

\author{Hong Jian Zhao}
\email{physzhaohj@jlu.edu.cn}
\affiliation{Key Laboratory of Material Simulation Methods and Software of Ministry of Education, College of Physics, Jilin University, Changchun 130012, China}
\affiliation{Key Laboratory of Physics and Technology for Advanced Batteries (Ministry of Education), College of Physics, Jilin University, Changchun 130012, China}
\affiliation{International Center of Future Science, Jilin University, Changchun 130012, China}

\author{Yu Xie}
\email{xieyu@jlu.edu.cn}
\affiliation{Key Laboratory of Material Simulation Methods and Software of Ministry of Education, College of Physics, Jilin University, Changchun 130012, China}
\affiliation{Key Laboratory of Physics and Technology for Advanced Batteries (Ministry of Education), College of Physics, Jilin University, Changchun 130012, China}
\affiliation{International Center of Future Science, Jilin University, Changchun 130012, China}

\author{Laurent Bellaiche}
\affiliation{Smart Ferroic Materials Center, Physics Department and Institute for Nanoscience and Engineering, University of Arkansas, Fayetteville, Arkansas 72701, USA}
\affiliation{\mbox{Department of Materials Science and Engineering, Tel Aviv University, Ramat Aviv, Tel Aviv 6997801, Israel}}

\date{\today}

\begin{abstract}
While scandium-doped aluminum nitride (AlScN) exhibits robust ferroelectricity and excellent thermal stability, its utility is limited by an exceptionally high coercive field ($E_c$) for polarization switching. Unraveling the atomistic switching dynamics is therefore critical for tailoring $E_c$. Here, we combine density functional theory and machine-learning molecular dynamics to elucidate the polarization switching mechanisms in AlScN over various Sc concentrations and applied electric fields. We find that excessive lattice strain strictly prohibits collective polarization switching, but the pre-existing domain walls relieve strain and lead to a distinct switching dynamics---dictating a field-dependent switching mechanism. At low electric fields, switching occurs via gradual domain-wall propagation consistent with the Kolmogorov-Avrami-Ishibashi model. In contrast, high fields stimulate additional nucleation, driving a rapid, homogeneous reversal process described by the simultaneous non-linear nucleation and growth model. These findings highlight the critical role of domain-wall dynamics and suggest domain engineering as a viable strategy to tailor coercive fields in AlScN and related ferroelectrics.
\end{abstract}

\begin{bibunit}[apsrev4-2]

\maketitle

\textit{Introduction---}Aluminum scandium nitrides with varying Sc concentrations (i.e., $\text{Al}_{1-x}\text{Sc}_{x}\text{N}$) are promising III-V semiconductors with experimentally demonstrated ferroelectricity \cite{fichtner_2019}. Subsequent studies have shown exceptional properties in $\text{Al}_{1-x}\text{Sc}_{x}\text{N}$, including large remanent polarizations ($P_r \approx 165\,\mu\mathrm{C}/\mathrm{cm}^2$ \cite{wolff_2024}), nearly square polarization–electric field hysteresis loops \cite{fichtner_2019}, and thermal stability up to $1100\,^\circ\mathrm{C}$ \cite{lslam_2021}. These features are ideal for designing memory devices for harsh-environment applications~\cite{fichtner_2019,guido_2023}. Despite this, the exceptionally high coercive field ($E_c \sim \text{MV/cm}$) for polarization switching in $\text{Al}_{1-x}\text{Sc}_{x}\text{N}$ ferroelectrics---two to three orders of magnitude greater than conventional perovskite ferroelectrics \cite{wang_2020,hayashi_1993,pontes_2000}---severely limits their device endurance and energy efficiency. Discovering pathways to mitigate $E_c$ is therefore imperative, which relies on elucidating the atomistic polarization switching dynamics.

Two primary mechanisms of polarization switching are generally considered: the collective atomic displacements across a nonpolar intermediate state, and localized reversal governed by domain-wall (DW) nucleation and propagation. Previous density functional theory (DFT) studies present conflicting views on the switching mechanism. Early works suggested a collective polarization reversal behavior traveling via either a nonpolar hexagonal \cite{konishi_2016,farrer_2002} or a $\mathrm{\beta-BeO}$-like \cite{liu_2023} intermediate state. In contrast, other works proposed a localized, column-by-column nucleation \cite{krishnamoorthy_2021} or a composition-driven transition between collective and localized modes \cite{lee_2024}. This controversy may be induced by the limited cell size of standard DFT simulations that prevent the accurate quantification of extended domain-wall dynamics \cite{behrendt_2024}. Crucially, recent experiments have underscored the indispensable role of DWs in polarization reversal \cite{schonweger_2023, wolff_2025}. For instance, Sch\"onweger {\it et al.} employed scanning transmission electron microscopy to directly image inversion domain boundaries \cite{schonweger_2023}, and Wolff {\it et al.} revealed the atomically sharp structures of field-induced vertical DWs and residual N-polar spike domains in $\text{Al}_{1-x}\text{Sc}_{x}\text{N}$ \cite{wolff_2025}. These observations indicate that the polarization reversal is inherently a DW-mediated process. Thus, there is a pressing need to develop large-scale atomistic models capable of describing DWs to systematically elucidate how these boundaries govern the reversal process.

In this Letter, we investigate the polarization switching mechanism in ferroelectric $\text{Al}_{1-x}\text{Sc}_{x}\text{N}$ using DFT and machine-learning molecular dynamics (MD) simulations \cite{SM}\nocite{kresse_1996, blochl_1994, perdew_2008, xu_2019, nose_1984, hoover_1985, martyna_1992, patidar_2024, stefan_2019}. By introducing pre-existing 180$^{\circ}$ DWs as strain-relief pathways, we demonstrate that DWs enable low-energy polarization switching via two field-dependent mechanisms: (1) at low fields, switching proceeds via gradual domain-wall propagation; and (2) additional nucleation events occur, accelerating reversal at high fields. Crucially, DWs universally govern switching across Sc compositions (15-30\%), directly challenging prior theoretical predictions of a composition-dependent transition from collective to individual switching mechanism \cite{lee_2024}. Our simulations further reveal that DWs reduce strain along the polar axis compared to the monodomain system by inducing a low-energy domain-wall-driven pathway, providing atomistic justification for experimentally observed domain dynamics, and therefore yielding atomistic insights for domain engineering in nitride ferroelectrics.

\begin{figure}[htbp]
  \centering
  \includegraphics[width=0.48\textwidth]{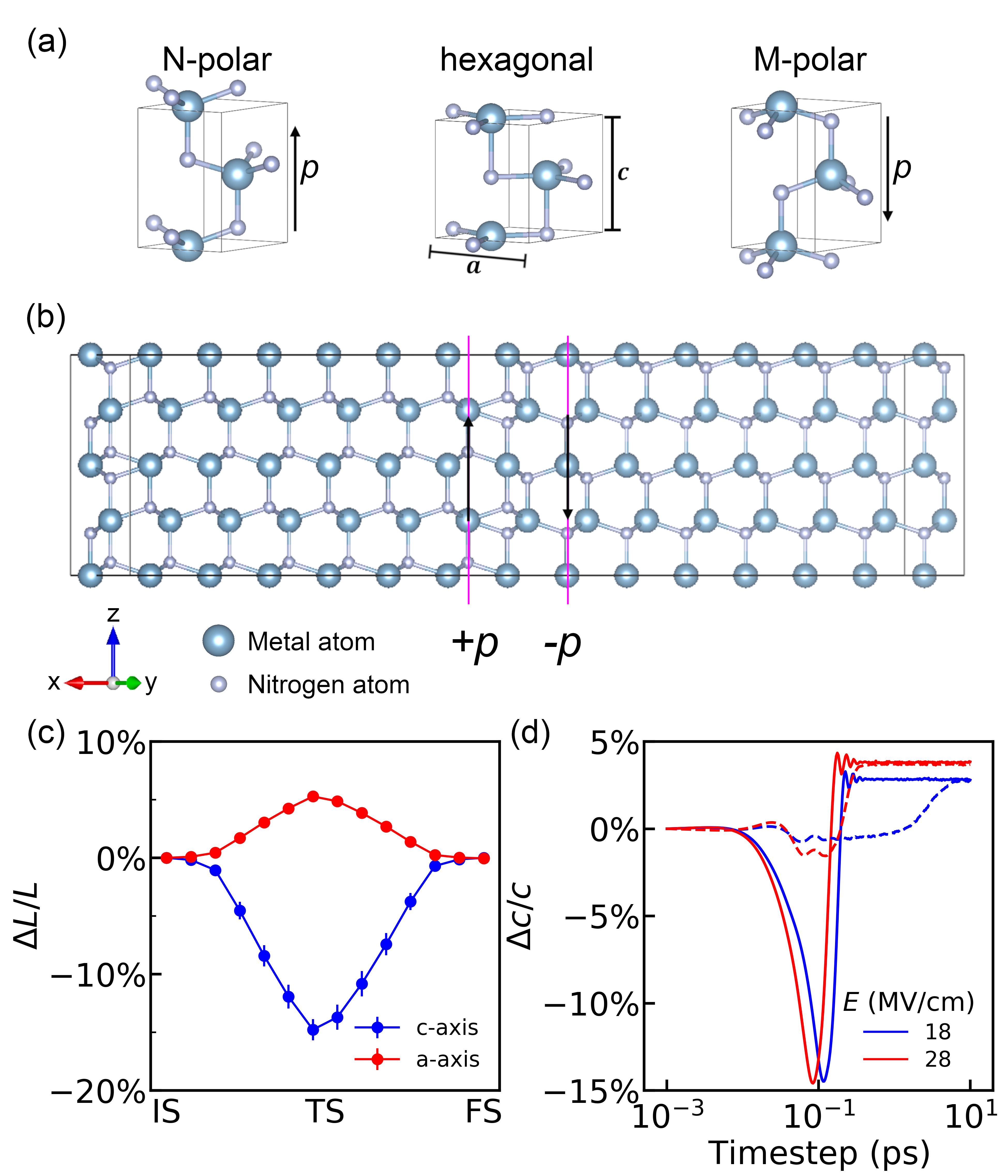}
  \caption{Structure, lattice response and strain evolution in $\mathrm{Al}_{1-x}\mathrm{Sc}_{x}\mathrm{N}$.~(a)~Collective path configurations: N-polar wurtzite (left), nonpolar hexagonal (middle) and M-polar wurtzite (right). Polarization lies along the $c$-axis. (b)~Schematic description of a multidomain configuration with N-polar (left) and M-polar (right) regions. 
  Dark blue and gray spheres represent metal (Al/Sc) and nitrogen (N) atoms, respectively. (c)~Evolution of the $a$- (red) and $c$-axis (blue) lattice strain in $\mathrm{Al}_{0.75}\mathrm{Sc}_{0.25}\mathrm{N}$ obtained from SS-NEB calculations. IS, TS and FS denote initial, transition and final states, respectively. Error bars indicate the standard deviation over multiple simulations.~(d)~Time evolution of the strain along the $c$-axis in $\mathrm{Al}_{0.75}\mathrm{Sc}_{0.25}\mathrm{N}$ under electric fields of 18 (blue) and 28 (red) MV/cm for monodomain (solid) and multidomain (dashed) systems.}
  \label{fig:fig1}
\end{figure}

\emph{Strain behavior}---We begin by assessing whether the classical collective switching mechanism is reasonable under realistic conditions. In the monodomain system, collective switching proceeds through a concerted displacement of cations and anions along the $c$-axis. The transition occurs between the N-polar phase with upward polarization and the M-polar phase with downward polarization, through the hexagonal transition state (see Fig.~\ref{fig:fig1}(a)). Reference \cite{lee_2024} proposes that collective switching governs polarization reversal in $\mathrm{Al}_{1-x}\mathrm{Sc}_x\mathrm{N}$ for Sc concentrations up to $x=0.28$, where massive lattice strain is required to traverse the nonpolar hexagonal transition state \cite{lee_2024,akiyama_2025,liu_2023}. To evaluate this, we perform solid-state nudged elastic band (SS-NEB) calculations \cite{henkelman_2000_0, henkelman_2000_1} on four 72-atom $\mathrm{Al}_{0.75}\mathrm{Sc}_{0.25}\mathrm{N}$ special quasirandom structure (SQS) supercells \cite{zunger_1990,a_2009}. As shown in Fig.~\ref{fig:fig1}(c), the $c$-axis strain reaches nearly 15\% during reversal, consistent with previous reports \cite{akiyama_2025}. However, such extreme strain is physically prohibitive in real films \cite{chen_2025}, undermining the practical viability of the collective switching pathway in $\text{Al}_{1-x}\text{Sc}_{x}\text{N}$.

Alternatively, experimental observations of domain boundaries point to a DW-mediated mechanism \cite{schonweger_2023}, which may circumvent prohibitive strain through localized distortions. To investigate this, we perform large-scale molecular dynamics (MD) simulations at 300~K in the $NPT$ ensemble across Sc concentrations of $x \in \{0.15, 0.20, 0.25, 0.30\}$ using a DFT-trained deep neural-network potential \cite{li_2026} (see Supplementary Material (SM) for training details \cite{SM}). To ensure a consistent number of switchable dipoles, we construct and compare a 15,360-atom uniform N-polar monodomain (see Fig.~\ref{fig:fig1}(a)) against a 30,720-atom multidomain system incorporating $180^\circ$ DWs (see Fig.~\ref{fig:fig1}(b)). To eliminate biases from local atomic arrangements, five independent unconstrained random substitution models are constructed for each concentration. We adopt this specific distribution because it is most favorable for realizing complete polarization switching at lower electric fields when compared against Sc-clustered and controlled-random structures (see SM \cite{SM}, Sec.~I~D). After a 10~ps relaxation, we evaluate the maximum $c$-axis strain under various applied fields $E$ over a 1~ns window. During polarization reversal, the monodomain system undergoes a severe $c$-axis compression that is determined by Sc concentration and is insensitive to $E$ (see SM \cite{SM}, Fig.~S7(a)), consistent with SS-NEB predictions. This dominant compression, followed by a strain recovery and a weaker, field-dependent elongation (see Fig.~\ref{fig:fig1}(d)), confirms a collective switching mechanism across all compositions. Remarkably, near the threshold field $E_s$ (defined as the minimum field for complete switching within 1~ns \cite{zhu_2025}), a distinct mode emerges for 15\% Sc. Characterized by reduced strain and a slower, multi-stage reversal with sustained lattice distortion (see SM \cite{SM}, Fig.~S7(b)), this alternative pathway indicates a transition from collective to distortion-assisted pathway under weak fields (see SM \cite{SM}, Sec.~II~A).

In contrast to the monodomain, the multidomain system exhibits substantially reduced $c$-axis strain throughout reversal (see SM \cite{SM}, Fig.~S7(a)). The observed maximum strain is driven by a post-reversal elongation that scales almost linearly with the applied field (see Fig.~\ref{fig:fig1}(d) and Fig.~S7(a) in SM \cite{SM}). Moreover, the strain evolution shows two distinct patterns: gradual elongation at low fields, and a slight compression followed by rapid elongation at high fields. These observations indicate that DWs mediate a localized switching mechanism by acting as stress-relief centers \cite{Jiang2024-AlNScN}, effectively mitigating local pressure accumulation. These findings highlight the critical role of DWs in the reversal pathway and motivate further investigation into their field-dependent dynamics.

\begin{figure}[htbp]
  \centering
  \includegraphics[width=0.48\textwidth]{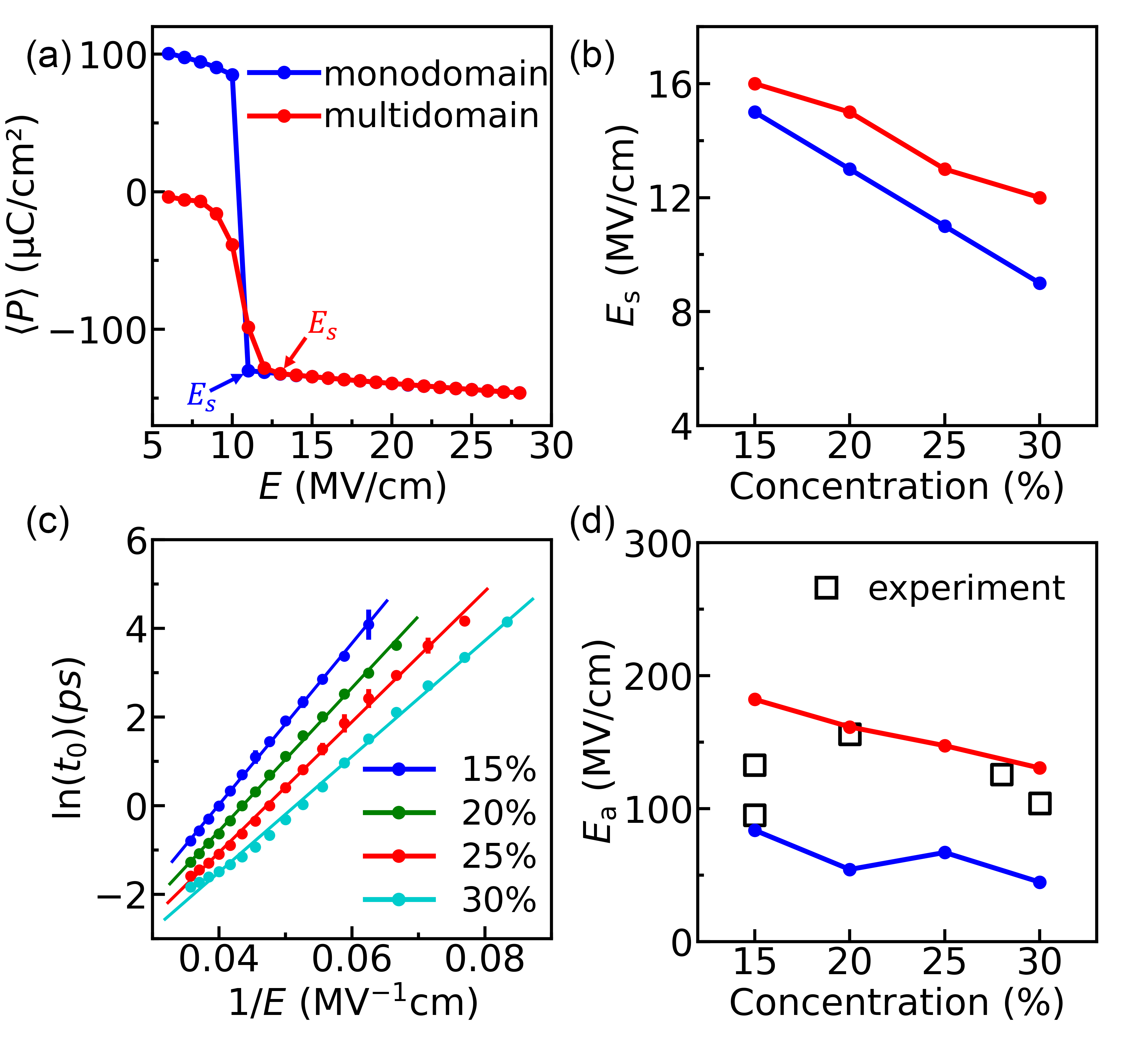}
  \caption{Polarization switching characteristics in $\mathrm{Al}_{1-x}\mathrm{Sc}_{x}\mathrm{N}$ under various electric fields and Sc concentrations. (a)~Field‐dependent polarization for $\mathrm{Al}_{0.75}\mathrm{Sc}_{0.25}\mathrm{N}$ in monodomain (blue) and multidomain (red) simulations.~(b)~Switching field $E_s$ as a function of Sc concentration.~(c)~$\ln(t_0)$ versus $1/E$ for multidomain system. Colors indicate different Sc concentrations.~(d)~Comparison of $E_a$ extracted from monodomain, multidomain, and experiments \cite{gao_2024,lu_2024,yasuoka_2023,guido_2024,guido_2024-1}. For the monodomain system showing multiple linear Merz regimes (see Fig.~S8(b) \cite{SM}), only $E_a$ closest to experiments is shown.}
  \label{fig:fig2}
\end{figure}

\emph{Polarization characteristics}---We next examine the field-dependent polar state after a simulation window of 1~ns. As shown in Fig.~\ref{fig:fig2}(a), the polarization response reveals two distinct electro-mechanical regimes. For the monodomain system, below $E_s$, the applied field compresses the lattice along the polar direction, driving a nearly linear reduction in polarization without triggering local switching. Alternatively, above $E_s$, the post-switching lattice elongation slightly increases the reversed polarization, a manifestation of the converse piezoelectric effect. For the multidomain system, a slightly higher $E_s$ is accompanied by a non-linear sub-threshold polarization response. Consistent with our strain analysis (see Fig.~\ref{fig:fig1}(d)), this indicates that pre-existing DWs undergo slow propagation, enabling progressive, localized yet incomplete reversal under weak fields. From this strain-field dependence, we calculate the piezoelectric coefficient (see SM \cite{SM}, Fig.~S7(c)). The derived value is slightly higher than experimental measurements since the $NPT$ ensemble neglects the mechanical clamping and residual strain present in actual thin films.

As summarized in Fig.~\ref{fig:fig2}(b), $E_s$ decreases with increasing Sc content for both systems, consistent with the experimentally observed reduction of coercive field upon Sc alloying \cite{fichtner_2019}. The switching kinetics follow the Merz's law, $t_0 \propto \exp(E_a/E)$ \cite{merz_1954}. This is evidenced by the linear scaling of $\ln t_0$ versus $1/E$ (see Fig.~\ref{fig:fig2}(c) for the multidomain system and Fig.~S8(b) for all systems \cite{SM}), where $t_0$ is the characteristic switching time and $E_a$ is the temperature-dependent activation field. The extracted $E_a$ values, which parameterize the macroscopic switching barrier, consistently decrease at higher Sc concentrations (see Fig.~\ref{fig:fig2}(d)). Notably, the close agreement of $E_a$ between multidomain system and experiments \cite{gao_2024,lu_2024,yasuoka_2023} validates our atomistic model and underscores the critical role of DWs in polarization switching.

\begin{figure}[htbp]
  \centering
  \includegraphics[width=0.48\textwidth]{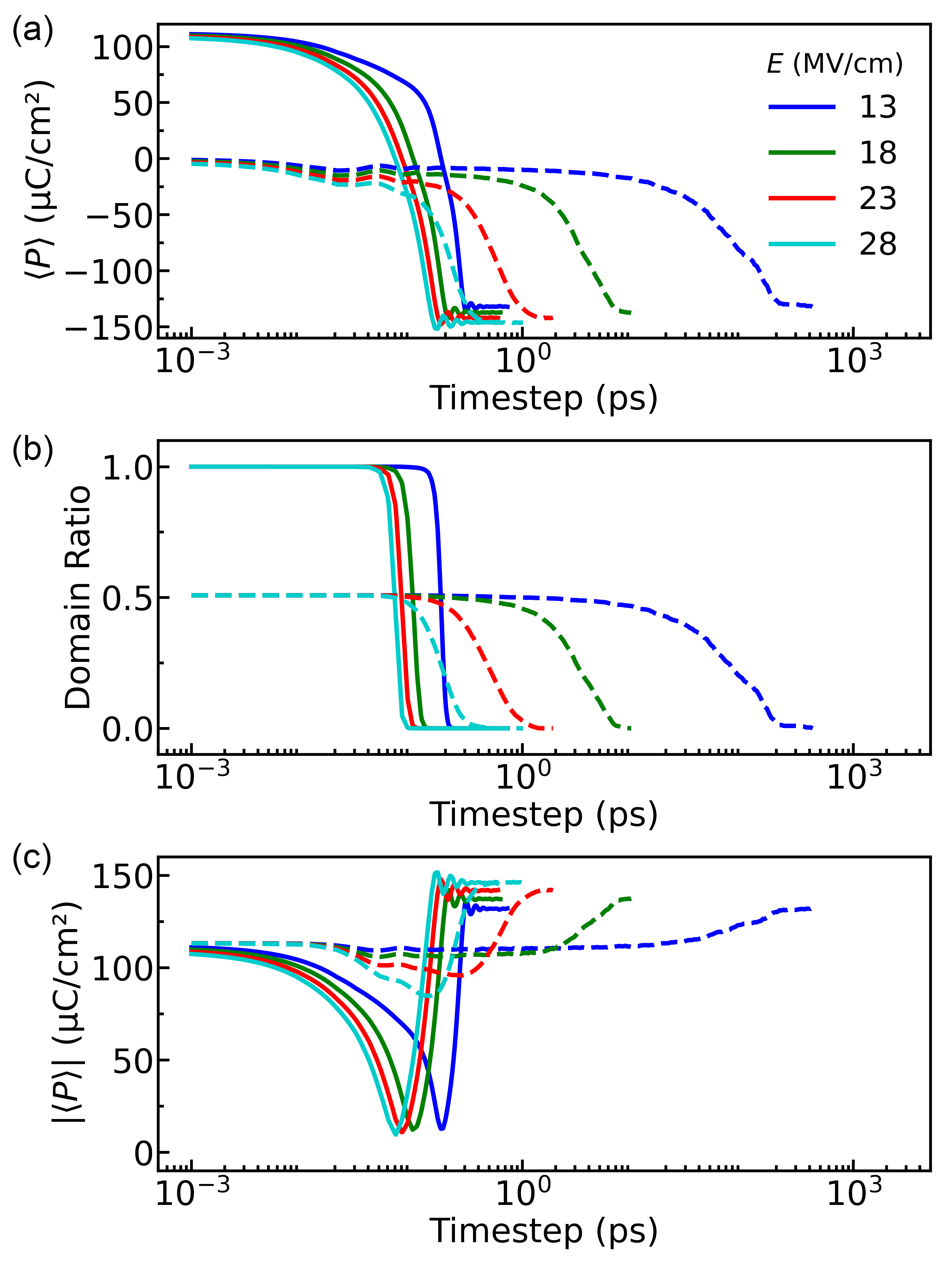}
  \caption{Temporal evolution of polarization and domain metrics of $\mathrm{Al}_{0.75}\mathrm{Sc}_{0.25}\mathrm{N}$ under different switching fields.~Time evolution of (a) the average polarization $\langle P \rangle$, (b) the domain ratio, and (c) the average local polarization magnitude $\langle |P| \rangle$ under $E$ = 13 (blue), 18 (green), 23 (red), and 28 (cyan) MV/cm. Solid and dashed lines indicate monodomain and multidomain systems.}
  \label{fig:fig3}
\end{figure}

\emph{Switching dynamics}---Moving beyond steady-state endpoints, we examine the time-resolved polarization and reversed domain ratio under representative fields ($E = 13, 18, 23,$ and $28$ MV/cm). The domain ratio is defined as the proportion of the total volume occupied by domains with a positive polarization orientation. As an example, the monodomain $\mathrm{Al}_{0.75}\mathrm{Sc}_{0.25}\mathrm{N}$ displays a two-stage evolution (see Figs.~\ref{fig:fig3}(a, b)). Initially, the polarization decreases gradually while the domain ratio remains unchanged, characterizing an elastic regime dominated by compressive lattice strain. Subsequently, a rapid lattice elongation drives an abrupt reversal in both polarization and domain ratio. The temporal evolution of the average local polarization magnitude ($\langle |P| \rangle$) also undergoes a sharp initial drop and subsequent recovery (see Fig.~\ref{fig:fig3}(c)). For all applied fields, the alignment of these minima indicates a highly homogeneous switching process, characteristic of a uniform collective reversal. This behavior is robustly observed across other Sc concentrations (see SM \cite{SM}, Figs.~S13(a, b)), confirming the general validity of this switching pathway.

In contrast, the multidomain system exhibits a field-dependent switching mechanism. Across all applied fields, both polarization and domain ratios decrease gradually and synchronously (see Figs.~\ref{fig:fig3}(a, b)). However, the temporal evolution of $\langle |P| \rangle$ bifurcates depending on the field strength (see Fig.~\ref{fig:fig3}(c)). At low fields (13 and 18 MV/cm), this magnitude remains stable throughout the reversal. Conversely, at high fields (23 and 28 MV/cm), a pronounced transient valley emerges. Notably, these local minima of $\langle |P| \rangle$ remain significantly higher than the monodomain values, indicating a highly inhomogeneous and localized switching process. Consistently observed across all Sc concentrations (see SM \cite{SM}, Figs.~S13(c, d)), these dynamics confirm that DWs dictate a localized reversal pathway distinct from collective switching.

To further analyze the polarization switching behavior, we fit the time-dependent polarization curves using three models: the Kolmogorov-Avrami-Ishibashi (KAI) model \cite{avrami_1940,ishibashi_1971,kolmogorov_1937}, the nucleation-limited-switching (NLS) model \cite{tagantsev_2002} and the simultaneous non-linear nucleation and growth (SNNG) model \cite{yazawa_2023}. As shown in Figs.~S14 and S15 of SM \cite{SM}, the KAI model can well describe the polarization dynamics of $\text{Al}_{1-x}\text{Sc}_{x}\text{N}$ ($x = 0.15, 0.2, 0.25, 0.3$) under low electric fields; at higher fields, the SNNG model yields the lowest loss, consistent with its suitability for describing abrupt reversal associated with new nucleation events. Furthermore, the extracted Avrami exponents for the KAI model fall between 1 and 4 (see Eq.~(S4) and Fig.~S16(a) of the SM \cite{SM}). We recall that a recent experiment reports the KAI switching kinetics in $\text{Al}_{0.8}\text{Sc}_{0.2}\text{N}$ with an Avrami exponent of $\approx 2$ \cite{yasuoka_2023}. Overall, these results demonstrate that classical kinetic models, particularly the KAI, remain effective in capturing the polarization dynamics of multidomain $\text{Al}_{1-x}\text{Sc}_{x}\text{N}$, while the SNNG model provides additional physical insight into high-field abrupt switching. 

\begin{figure}[htbp]
  \centering
  \includegraphics[width=0.48\textwidth]{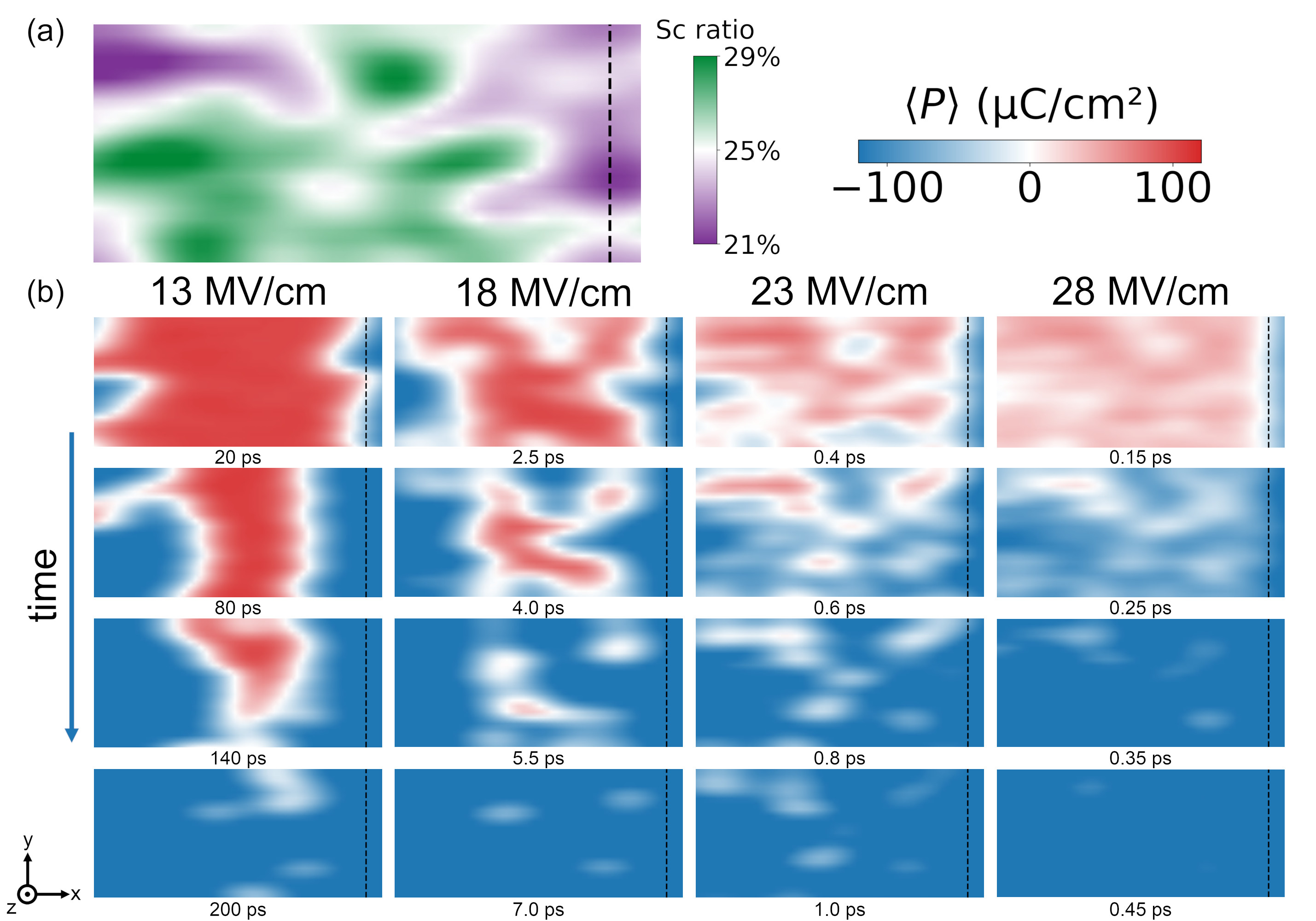}
  \caption{Evolution of polarization patterns in the $xy$-plane of $\mathrm{Al}_{0.75}\mathrm{Sc}_{0.25}\mathrm{N}$.~(a)~The spatial distribution of the local Sc concentration projected onto the $xy$-plane.~(b)~The evolution under different electric fields $E$ = 13, 18, 23 and 28 MV/cm. Only the left half of the system is displayed. The color bar in the top-right indicates the magnitude of the local polarization.}
  \label{fig:fig4}
\end{figure}

To elucidate the spatial switching dynamics, we project the time-dependent polarization patterns onto the $x$-$y$ plane (see Fig.~\ref{fig:fig4}(b)). At low fields (13 and 18 MV/cm), reversal proceeds primarily via the lateral expansion of pre-existing 180$^\circ$ DWs. Atomistically, the four-fold coordination fractions ($f_{\text{AlN}_4}$ and $f_{\text{ScN}_4}$) exhibit only minor fluctuations, confirming that severe structural distortions are strictly localized at the DWs. Notably, these advancing DWs exhibit a zigzag morphology, consistent with recent experimental observations \cite{huang_2026}. At high fields (23 and 28 MV/cm), the mechanism shifts to a nucleation-dominated regime. This transition is accompanied by a pronounced transient dip in both $f_{\text{AlN}_4}$ and $f_{\text{ScN}_4}$ (see SM \cite{SM}, Fig.~S10), signifying the widespread bond-breaking events central to the SNNG model. While these nucleation events spatially correlate with local Sc concentration (Fig.~\ref{fig:fig4}(a)), the global kinetic pathway is dictated by the field strength. These observations establish a field-driven transition from localized DW motion to widespread nucleation.

To further determine the critical field for this mechanistic crossover, we introduce an energetic descriptor $\Delta V_{max}$ (the maximum potential energy difference along the switching trajectory). $\Delta V_{max}$ is highly sensitive to the reversal mode: localized DW propagation entails minor energy variations, whereas widespread nucleation induces a pronounced energy peak (see SM \cite{SM} for details). Evaluating this field-dependent signature reveals critical transition fields between 19 and 27 MV/cm across the studied Sc concentrations (see Table~S1 in SM \cite{SM}).

\emph{Conclusion}---In summary, our investigations combining both SS-NEB calculations and ML-MD simulations establish switching mechanisms in $\mathrm{Al}_{1-x}\mathrm{Sc}_x\mathrm{N}$ for Sc concentrations of 15-30\%. While collective switching induces large lattice strains, multidomain configuration with pre-existing DWs enables strain-relieved switching pathways. Simulations under various $E$ further uncover the interplay between field strength and switching mechanism. These results provide microscopic insights into domain-wall dynamics in nitride ferroelectrics and emphasize the importance of domain engineering in tuning coercive fields and improving device performance.

Recently, Huang~\textit{et al.}~utilized the NLS model to describe the device-scale polarization switching kinetics in $\text{Al}_{0.75}\text{Sc}_{0.25}\text{N}$ \cite{huang_2026}. As shown in the End Matter section, this macroscopic NLS behavior can be understood by our atomistic domain-wall-driven switching mechanism. Note also that, experimentally, the NLS and KAI/SNNG kinetics can be distinguished by the field dependence of the full width at half maximum (FWHM) extracted from the polarization switching measurements (see SM \cite{SM}, Eq.~(S3)). A linear dependence of the FWHM on $1/E^2$ indicates NLS behavior, whereas a narrow, field-independent FWHM supports the KAI/SNNG framework \cite{pirro_2022, yang_2026, lu_2024, yasuoka_2023, guido_2024, guido_2025}.

\emph{Acknowledgments}---This work was supported by the National Key R\&D Program of China (Grant No. 2023YFB3003001) and the National Natural Science Foundation of China (Grants Nos. 12274174). Work at the University of Arkansas was supported by the U.S. Department of Energy (DOE), Office of Science, Basic Energy Sciences (BES) under Award No. DE-SC0025479. We also thank the Changchun Computing Center for providing comprehensive computing resources and technical support throughout the completion of this work.

\nocite{lee_2024_1, jo_2007, wei_2021}

\putbib[reference]

\onecolumngrid
\begin{center}
    \vspace{20pt}
    \textbf{\large End Matter}
    \vspace{10pt}
\end{center}

\twocolumngrid

In this section, we discuss our idealized MD simulations on $\text{Al}_{1-x}\text{Sc}_{x}\text{N}$ in the context of experimental device-scale observations. It is crucial to recognize that macroscopic switching kinetics are a convolution of intrinsic lattice responses and extrinsic microstructural heterogeneities \cite{yang_2026, wolff_2025}. By simulating an ideal supercell, we explicitly isolate the fundamental reversal mechanism from these extrinsic effects. Our results demonstrate that the intrinsic DW propagation follows the KAI and SNNG frameworks. This clean limit is supported by recent experiments observing both behaviors, where KAI kinetics dominate at lower fields and transition to the SNNG regime under high electric fields \cite{yasuoka_2023, guido_2025, guido_2024-1, guido_2024, yazawa_2023}.

Importantly, this DW mechanism could serve as the physical foundation for the observed NLS behavior \cite{huang_2026, gao_2024, lee_2024_1}. Within a unified theoretical framework \cite{jo_2007, tagantsev_2002}, KAI and NLS describe similar underlying DW motions but considering different statistical distributions of switching times. In our defect-free simulations, DW propagation yields a single characteristic time (Dirac delta distribution), manifesting as KAI kinetics \cite{jo_2007, lu_2024, wei_2021}. In practical films, however, local pinning sites impede this advancing DW \cite{guido_2024-1, guido_2024, gao_2024}. This restriction drastically broadens the switching time into a Lorentzian distribution, transforming the macroscopic kinetics into NLS behavior \cite{jo_2007}. The fact that our intrinsic simulations reproduce the zigzag DW morphology observed in NLS-dominated devices further solidifies this connection \cite{huang_2026}. Therefore, these findings demonstrate that in $\text{Al}_{1-x}\text{Sc}_{x}\text{N}$ the macroscopic NLS behavior is fundamentally governed by the atomistic domain-wall propagation mechanism, effectively bridging device-scale observations with their microscopic origins.

\end{bibunit}
\onecolumngrid

\newpage
\clearpage

\title{\Large\bfseries Supplementary Material of\\[0.3em]
``Domain-Wall Mediated Polarization Switching in Ferroelectric AlScN:\\[0.3em]
Strain Relief and Field-Dependent Dynamics''}

\author{Xiangyu Zheng}
\affiliation{Key Laboratory of Material Simulation Methods and Software of Ministry of Education, College of Physics, Jilin University, Changchun 130012, China}

\author{Charles Paillard}
\email{paillard@uark.edu}
\affiliation{Smart Ferroic Materials Center, Physics Department and Institute for Nanoscience and Engineering, University of Arkansas, Fayetteville, Arkansas 72701, USA}

\author{Dawei Wang}
\affiliation{Smart Ferroic Materials Center, Physics Department and Institute for Nanoscience and Engineering, University of Arkansas, Fayetteville, Arkansas 72701, USA}

\author{Peng Chen}
\affiliation{Guangdong Technion – Israel Institute of Technology, Guangdong 515063, China}
\affiliation{Department of Physics, Technion – Israel Institute of Technology, 32000 Haifa, Israel}

\author{Hong Jian Zhao}
\email{physzhaohj@jlu.edu.cn}
\affiliation{Key Laboratory of Material Simulation Methods and Software of Ministry of Education, College of Physics, Jilin University, Changchun 130012, China}
\affiliation{Key Laboratory of Physics and Technology for Advanced Batteries (Ministry of Education), College of Physics, Jilin University, Changchun 130012, China}
\affiliation{International Center of Future Science, Jilin University, Changchun 130012, China}

\author{Yu Xie}
\email{xieyu@jlu.edu.cn}
\affiliation{Key Laboratory of Material Simulation Methods and Software of Ministry of Education, College of Physics, Jilin University, Changchun 130012, China}
\affiliation{Key Laboratory of Physics and Technology for Advanced Batteries (Ministry of Education), College of Physics, Jilin University, Changchun 130012, China}
\affiliation{International Center of Future Science, Jilin University, Changchun 130012, China}

\author{Laurent Bellaiche}
\affiliation{Smart Ferroic Materials Center, Physics Department and Institute for Nanoscience and Engineering, University of Arkansas, Fayetteville, Arkansas 72701, USA}
\affiliation{\mbox{Department of Materials Science and Engineering, Tel Aviv University, Ramat Aviv, Tel Aviv 6997801, Israel}}

\date{\today}

\newcommand{\tabincell}[2]{\begin{tabular}{@{}#1@{}}#2\end{tabular}}
\renewcommand{\thefigure}{S\arabic{figure}}
\setcounter{figure}{0}
\renewcommand{\theequation}{S\arabic{equation}}
\setcounter{secnumdepth}{3}

\begin{bibunit}[apsrev4-2]

\maketitle
\onecolumngrid

\section{COMPUTATIONAL METHODS}

\subsection{Machine learning potential construction}

All first-principles calculations were performed using the Vienna \textit{Ab initio} Simulation Package (VASP) \cite{kresse_1996} with the projector augmented-wave (PAW) \cite{kresse_1996, blochl_1994} method and the Perdew-Burke-Ernzerhof exchange-correlation functional revised for solids (PBEsol) \cite{perdew_2008}. The plane-wave energy cutoff was set to 520 eV. Brillouin zone integration was carried out using a k-point spacing of 0.2 \AA$^{-1}$. The electronic self-consistent field (SCF) loop was converged to an energy threshold of $10^{-5}$ eV.

The training of the machine learning potentials, termed the attention coupled neural network (ACNN) \cite{li_2026}, began with the construction of an initial dataset using $\Gamma$-point density-functional theory (DFT) molecular dynamics (MD) simulations. These simulations were performed at 300 K, 700 K, and 1000 K for various configurations, including wurtzite structures with 25\% Sc content (with and without 180$^\circ$ domain walls (DWs)), as well as nonpolar rocksalt and zinc blende phases. Each system was simulated for up to 2000 MD steps, and 50 representative snapshots were selected from each trajectory and recomputed using high-precision SCF calculations. To further enhance the structural diversity, the dataset also included surface-terminated wurtzite structures along the $c$-axis and configurations with Sc clustering. Additionally, special quasirandom structures (SQS) \cite{zunger_1990} were generated using the Alloy Theoretic Automated Toolkit (ATAT) package \cite{a_2009} for the ferroelectric wurtzite phase at 12.5\%, 25\% and 37.5\% Sc contents, along with their corresponding DW configurations.

The ACNN model represented atomic interactions through elemental embedding and fitting networks. The embedding network adopted a two-layer architecture with dimensions $[64, 64]$, while the fitting network consisted of two layers with $[256, 256]$. The local atomic environment was expanded using radial basis functions of order $n_{\max}^r = 12$ and angular basis functions of order $n_{\max}^a = 10$. Both the radial cutoff $r_c^r$ and angular cutoff $r_c^a$ were set to 6.5 $\mathrm{\AA}$. The atomic energy contributions were predicted by the fitting network, and the total energy of the system was obtained as the sum of atomic energies. The framework was constructed to preserve essential physical symmetries: translational, rotational, and permutational invariance for scalar quantities such as the total energy, and rotational equivariance for tensorial quantities including atomic forces and stress tensors.

To effectively sample configurations relevant to polarization switching, an active learning approach was employed starting from an initial dataset of monodomain and multidomain configurations containing 256 atoms. The training set was expanded by iteratively incorporating configurations with high model uncertainty, sampled under equilibrium MD at 300 K, 700 K, and 1000 K as well as during polarization switching triggered by applied electric fields. This sampling covered Sc concentrations ranging from 6.25\% to 37.5\%. Through this iterative process, the final training dataset was enriched to a total of 5,320 structures, ensuring a robust and comprehensive model.

\subsection{Assessment of model applicability}

\begin{figure}[htbp]
  \centering
  \includegraphics[width=0.8\textwidth]{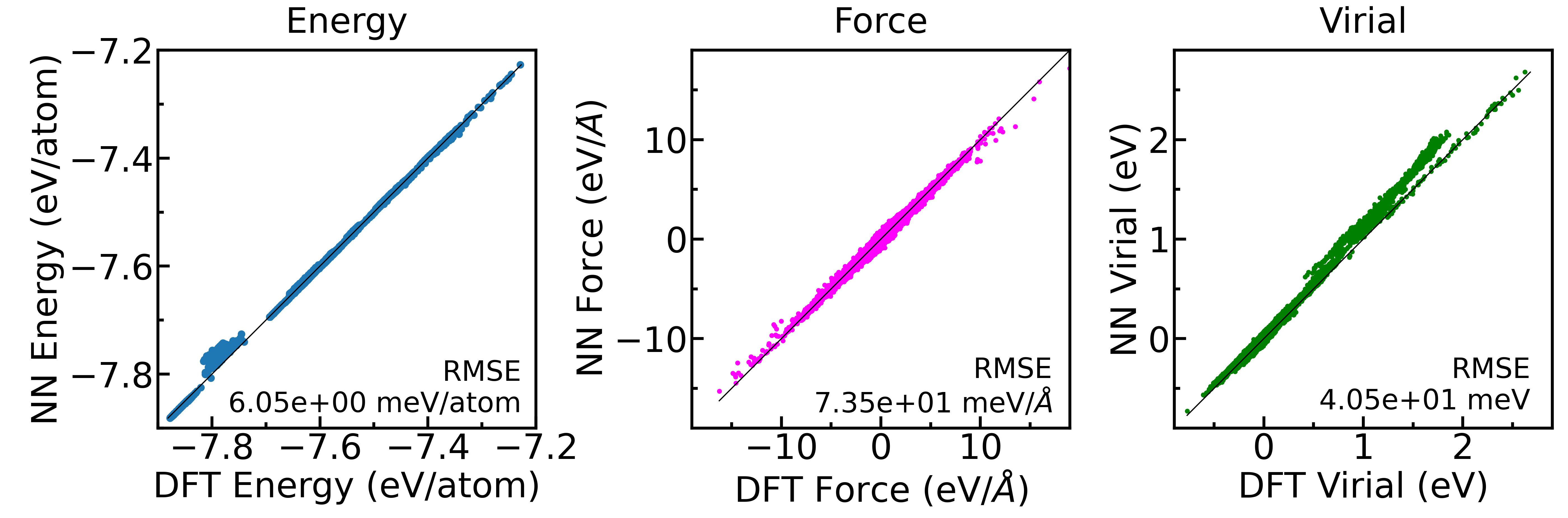}
  \caption{Comparison between ACNN predictions and DFT reference values across the training dataset. From left to right: total energy per atom, atomic force components, and virial stress tensor components.}
  \label{fig:eval}
\end{figure}

The accuracy and transferability of the trained machine learning potential were assessed through three complementary benchmarks. First, the model's predictive performance was evaluated against the full training set. As shown in Fig.~\ref{fig:eval}, the predicted values exhibited excellent agreement with DFT references across both energy and force components, confirming the model’s reliability within the sampled configuration space.

\begin{figure}[htbp]
    \centering
    \includegraphics[width=0.6\textwidth]{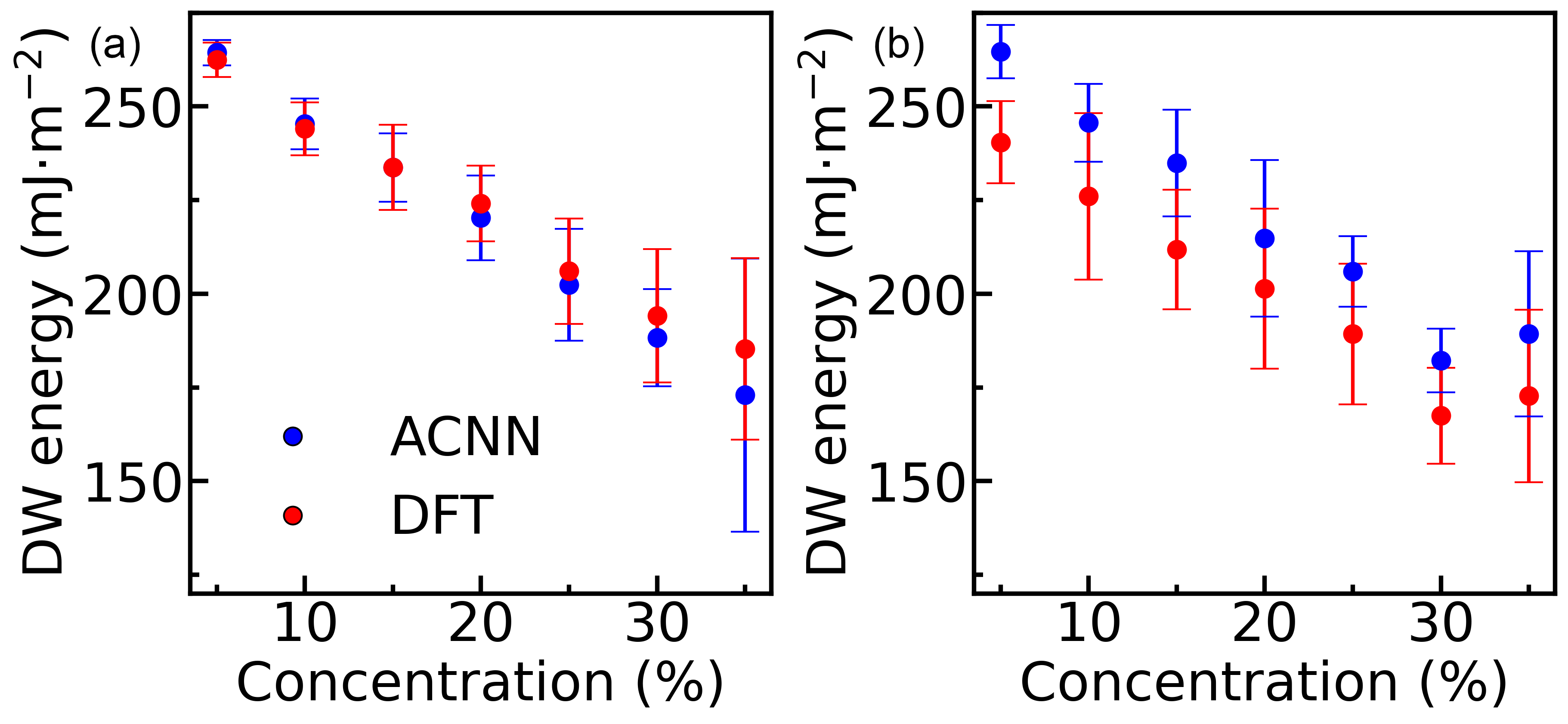}
    \caption{Comparison of DW energies between DFT and ACNN for $\mathrm{Al}_{1-x}\mathrm{Sc}_{x}\mathrm{N}$ with varying Sc concentrations (5–35\%).~(a)~DW energies computed using ACNN-relaxed structures;~(b)~DW energies computed using DFT-relaxed structures. In both panels, blue symbols represent ACNN predictions and red symbols denote DFT calculations. Error bars indicate statistical variations across different configurations at the same composition.}
    \label{fig:acnn_dft}
\end{figure}

In addition, we benchmarked the potential on a series of DW structures at different Sc concentrations (5–35\%). For each concentration, multiple DW configurations were generated. These structures were relaxed using the ACNN potential, and their DW energies were subsequently evaluated by both ACNN and DFT on the same optimized geometries. As illustrated in Fig.~\ref{fig:acnn_dft}(a), the ACNN predictions showed excellent agreement with the DFT results, indicating that the model performs reliably when the structural configurations remain within its applicability domain.

We further tested the comparison using DFT-relaxed DW structures at the same concentrations. As shown in Fig.~\ref{fig:acnn_dft}(b), although these structures were outside the training distribution, the ACNN predictions still reproduced the overall concentration dependence and achieved reasonable quantitative accuracy. This demonstrated that the potential possesses a certain degree of extrapolative capability.

\subsection{Molecular dynamics simulations}

All MD simulations were performed using the Ab initio atomic mateRial modEling Software (ARES) package \cite{xu_2019} with ACNN. The simulations were conducted in the isothermal-isobaric (NPT) ensemble at a fixed temperature of 300 K, maintained using a massive Nosé–Hoover chain thermostat \cite{nose_1984, hoover_1985, martyna_1992}. The pressure was maintained at 1 bar using a Nosé–Hoover barostat. The time step for integrating the equations of motion was set to 1 fs. Each simulation began with a 10 ps equilibration run to relax the structure under the given thermodynamic conditions. Following equilibration, an external electric field was applied along the $c$-axis, and its influence was included by modifying the atomic forces as $\mathbf{f}_i = z_i^* \cdot \mathbf{E}_{ext}$, where $\mathbf{f}_i$ is the interatomic force, $\mathbf{E}_{ext}$ is the external electric field applied along the polarization direction, and $z_i^*$ is the formal valence of the atom $i$. Only the effective charge component along the polarization axis was included, while transverse contributions were neglected. To explore field-induced switching behavior, we gradually extended the simulation time up to 1 ns by sequentially accumulating time blocks and monitoring the evolution of polarization. For statistical reliability, this procedure was repeated four times with different initial configurations.

Two types of supercells were adopted to represent distinct switching pathways. A wurtzite-type supercell, containing 15,360 atoms, was used to simulate monodomain systems, while a larger supercell, containing 30,720 atoms, was employed to capture multidomain behavior with pre-existing domain walls aligned along the $x$-axis. The polarization was computed from the relative displacements of cation–anion pairs along the $c$-axis. The total polarization at time $t$ is given by
\begin{equation}
  \mathbf{P}(t) = \sum_{i=1}^{N_{\text{cation}}} \mathbf{P}_{i}(t) = \sum_{i=1}^{N_{\text{cation}}}\frac{z_i^*}{V \cdot N_{\text{cation}}} \left( \mathbf{r}_{i}(t) - \frac{1}{3} \sum_{j=1}^{3} \mathbf{r}_{i,j}^{(N)}(t) \right),
\end{equation}
where $V$ is the volume of the simulation cell, $N_{\text{cation}}$ is the number of Al/Sc cations, $\mathbf{r}_i(t)$ denotes the position of the $i$th Al or Sc cation, and $\mathbf{r}_{i,j}^{(N)}(t)$ are the positions of its three nearest-neighbor nitrogen atoms at time $t$. The polarization is evaluated along the $c$-axis direction. The local polarization magnitude $\langle|\mathbf{P}|\rangle$ is defined as the average of the absolute values of these local dipoles, offering an approximate yet insightful descriptor of the polarization evolution during switching.

\subsection{Sensitivity of Switching Efficiency to Local Atomic Arrangement}

\begin{figure}[htbp]
  \centering
  \includegraphics[width=0.9\textwidth]{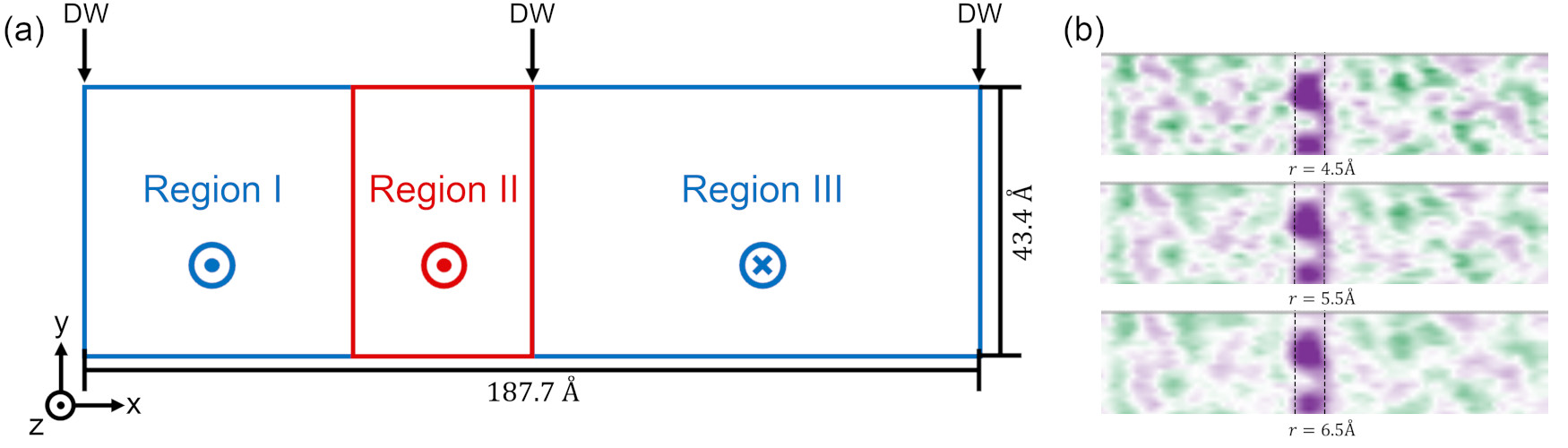}
  \caption{(a)~Schematic of the simulation setup. Region II denotes specific zone embedded between two random distribution regions (Region I and III). The distinct regions are initialized with different polarization orientations: upward for Regions I and II, and downward for Region III.~(b)~Comparison of local Sc concentration patterns calculated using different cutoff radii.}
  \label{fig:region}
\end{figure}

The main text utilizes unconstrained random substitution models to investigate the sensitivity of the switching mechanism to the local atomic arrangement of Sc. To further distinguish whether the switching efficiency is governed by the specific topology (clustering) or simply the local concentration, we constructed a partitioned supercell model containing 30,720 atoms, as illustrated in Fig.~\ref{fig:region}(a). This model was divided into three regions along the $x$-axis. Region II is defined as a specific zone (spanning 13~\AA, twice the cutoff radius) bordering the DW. Regions I and II were initialized with upward polarization ($+P_z$), and Region III with downward polarization ($-P_z$). This model explicitly positioned the DW at the interface between Region II and Region III.

We contrasted two distinct Sc distribution models within Region II (comprising 896 metal atoms): the Sc cluster model and the controlled random model. We defined two types of concentration: global and local. The global Sc concentration is defined as the ratio of Sc to total metal atoms in Region II. The local Sc concentration was calculated for each nitrogen (N) site by identifying its nearest-neighbor metal atoms within a specific cutoff radius. As illustrated in Fig.~\ref{fig:region}(b), we tested cutoff radii of 4.5, 5.5, and 6.5~{\AA} and observed negligible differences in the resulting patterns. Consequently, a uniform cutoff radius of 4.5~{\AA} was adopted to compute the local Sc concentration. In the Sc cluster model, we embedded varying numbers (2, 4, and 6) of non-overlapping spherical Sc clusters ($r=6.5$~\AA, centered on N sites) into Region II. These configurations yielded global Sc concentrations of approximately 11.4\%, 22.8\%, and 34.1\% within Region II, respectively. For the controlled random model, we constructed comparative models where Region II maintained the global Sc concentrations (11.4\%, 22.8\%, and 34.1\%) but with Sc atoms distributed via random substitution. To prevent accidental clustering in these random models, we imposed a geometric constraint ensuring that no Sc atom possesses a complete shell of Sc nearest neighbors. Regions I and III were then populated via unconstrained random substitution to maintain the global composition at $x=0.25$. For all configurations, simulations were performed under electric fields of 13, 18, 23, and 28 MV/cm. The simulation durations were set to 530 ps for 13 MV/cm and 20 ps for the higher fields, consistent with the total switching times observed in the unconstrained random $\mathrm{Al}_{0.75}\mathrm{Sc}_{0.25}\mathrm{N}$ alloy.

\begin{figure}[htbp]
  \centering
  \includegraphics[width=\textwidth]{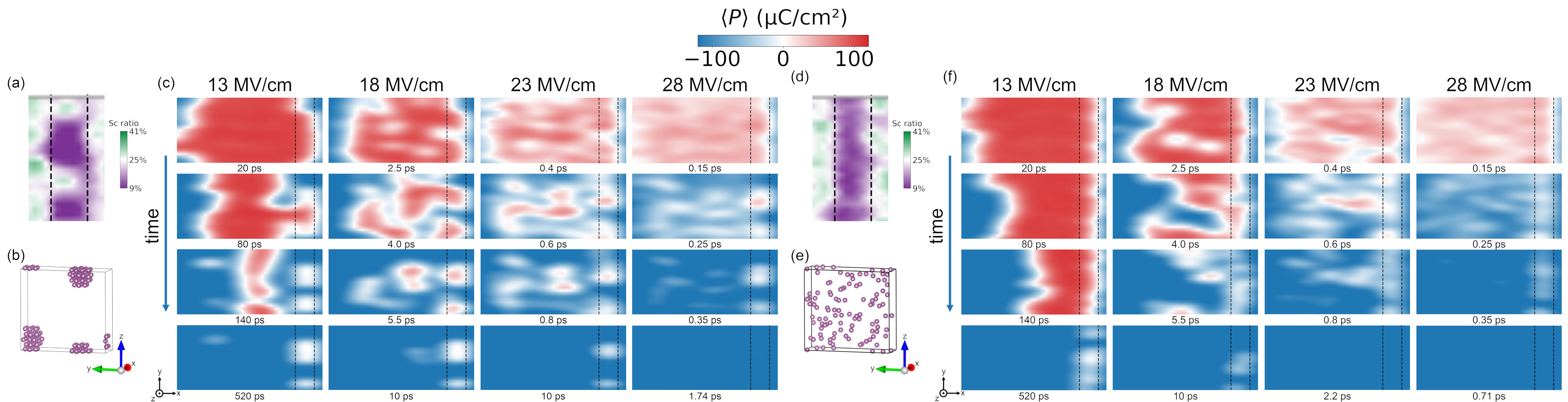}
  \caption{Comparison of polarization switching dynamics in $\mathrm{Al}_{0.75}\mathrm{Sc}_{0.25}\mathrm{N}$ between 2-cluster model (a--c) and the controlled random model with 11.4\% global Sc concentration (d--f). Structural characterization of Region II is shown in (a, d) local Sc concentration maps projected onto the $xy$-plane and (b, e) corresponding 3D atomic configurations showing Sc atoms (purple spheres).~(c,f)~The evolution of polarization patterns under different electric fields $E$ = 13, 18, 23, and 28 MV/cm. Only the left half of the system is displayed. Time proceeds from top to bottom. The color bar in the top indicates the magnitude of the local polarization. The vertical dashed lines mark Region II.}
  \label{fig:random-circle-2}
\end{figure}

\begin{figure}[htbp]
  \centering
  \includegraphics[width=\textwidth]{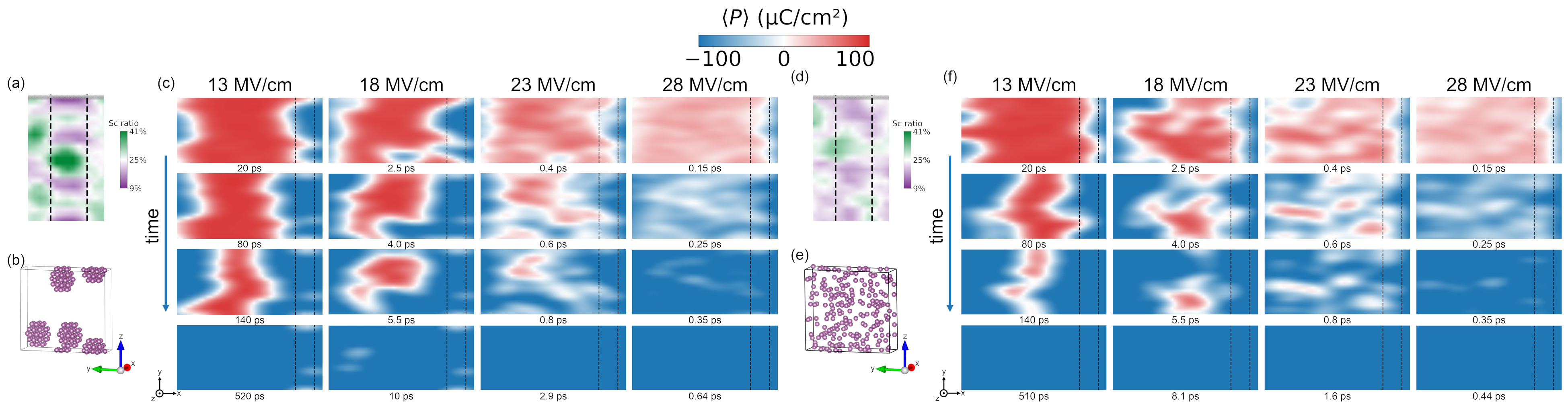}
  \caption{Comparison of polarization switching dynamics in $\mathrm{Al}_{0.75}\mathrm{Sc}_{0.25}\mathrm{N}$ between 4-cluster model (a--c) and the controlled random model with 22.8\% global Sc concentration (d--f). Structural characterization of Region II is shown in (a, d) local Sc concentration maps projected onto the $xy$-plane and (b, e) corresponding 3D atomic configurations showing Sc atoms (purple spheres).~(c,f)~The evolution of polarization patterns under different electric fields $E$ = 13, 18, 23, and 28 MV/cm. Only the left half of the system is displayed. Time proceeds from top to bottom. The color bar in the top indicates the magnitude of the local polarization. The vertical dashed lines mark Region II.}
  \label{fig:random-circle-4}
\end{figure}

\begin{figure}[htbp]
  \centering
  \includegraphics[width=\textwidth]{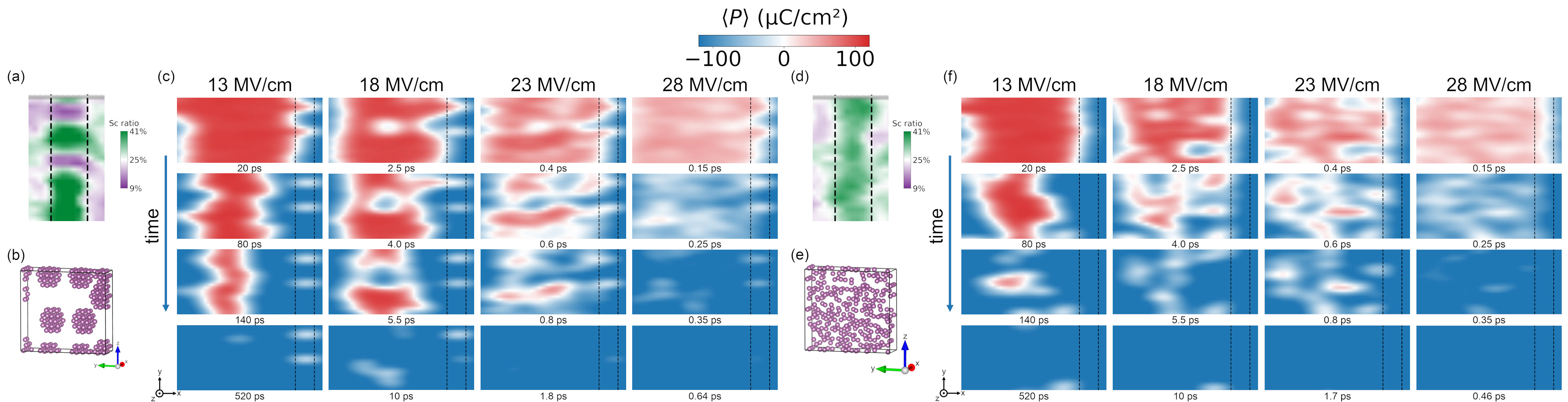}
  \caption{Comparison of polarization switching dynamics in $\mathrm{Al}_{0.75}\mathrm{Sc}_{0.25}\mathrm{N}$ between 6-cluster model (a--c) and the controlled random model with 34.1\% global Sc concentration (d--f). Structural characterization of Region II is shown in (a, d) local Sc concentration maps projected onto the $xy$-plane and (b, e) corresponding 3D atomic configurations showing Sc atoms (purple spheres).~(c,f)~The evolution of polarization patterns under different electric fields $E$ = 13, 18, 23, and 28 MV/cm. Only the left half of the system is displayed. Time proceeds from top to bottom. The color bar in the top indicates the magnitude of the local polarization. The vertical dashed lines mark Region II.}
  \label{fig:random-circle-6}
\end{figure}

The simulation results for the Sc cluster configurations and the controlled random models are shown in Figs.~\ref{fig:random-circle-2}-\ref{fig:random-circle-6}. Panels (a, d) map the local Sc concentration. The 2-cluster model (Fig.~\ref{fig:random-circle-2}(b)) exhibited non-uniform Sc distributions, where clustered Sc atoms in Region II acted as preferential nucleation sites at 13 MV/cm. However, this aggregation depleted Sc from the surrounding areas (down to $\approx$ 12\%), causing them to remain in an unswitched state. In contrast, in the controlled random model with 11.4\% global Sc concentration (Fig.~\ref{fig:random-circle-2}(e)), the 13 MV/cm field was insufficient to drive propagation in Region II due to the uniform local Sc concentration ($\approx$ 18\%). When the field was increased to 18 MV/cm, the controlled random model exhibited clear domain-wall propagation. By comparison, the 2-cluster model still retained unswitched regions, although they were reduced in area. This confirmed that the unswitched regions were not pinned but simply required a higher electric field to induce switching.

When the cluster number increased to 4 and 6 (raising global Sc concentration to 22.8\% and 34.1\%), the behavioral discrepancy between the models became less pronounced (see Figs.~\ref{fig:random-circle-4}-\ref{fig:random-circle-6}). Under these conditions, the controlled random models possessed sufficient local Sc concentration, exhibiting a similar zigzag wavefront morphology to that observed in the clustered models. However, unlike the random models, the 4- and 6-cluster models retained unswitched regions in Region II at 13 and 18 MV/cm, mirroring the behavior observed in the 2-cluster case.

Furthermore, a consistent mechanism transition was observed across all models: shifting from domain-wall propagation at lower fields ($13, 18$ MV/cm) to nucleation-dominated reversal at higher fields ($23, 28$ MV/cm). Mechanistically, this was supported by our bond analysis, which revealed that the minimum of $f_{\mathrm{ScN}_4}$ preceded that of $f_{\mathrm{AlN}_4}$ under the same conditions (Fig.~\ref{fig:cn4}). This temporal offset confirmed that Sc atoms act as preferential nucleation sites and drive the reversal of the surrounding atoms.

In summary, our supplementary simulations confirm that the switching process is fundamentally dictated by the local Sc concentration. While Sc clusters acted as preferential nucleation sites, their aggregation left the surrounding matrix depleted of Sc, thereby making switching more difficult. Consequently, the unconstrained random distribution proved to be more favorable for complete polarization switching at lower electric fields, as it avoided these pronounced heterogeneities.

\section{Polarization Switching Pathways}

\subsection{Lattice evolution under different switching conditions}

\begin{figure}[htbp]
    \centering
    \includegraphics[width=0.9\textwidth]{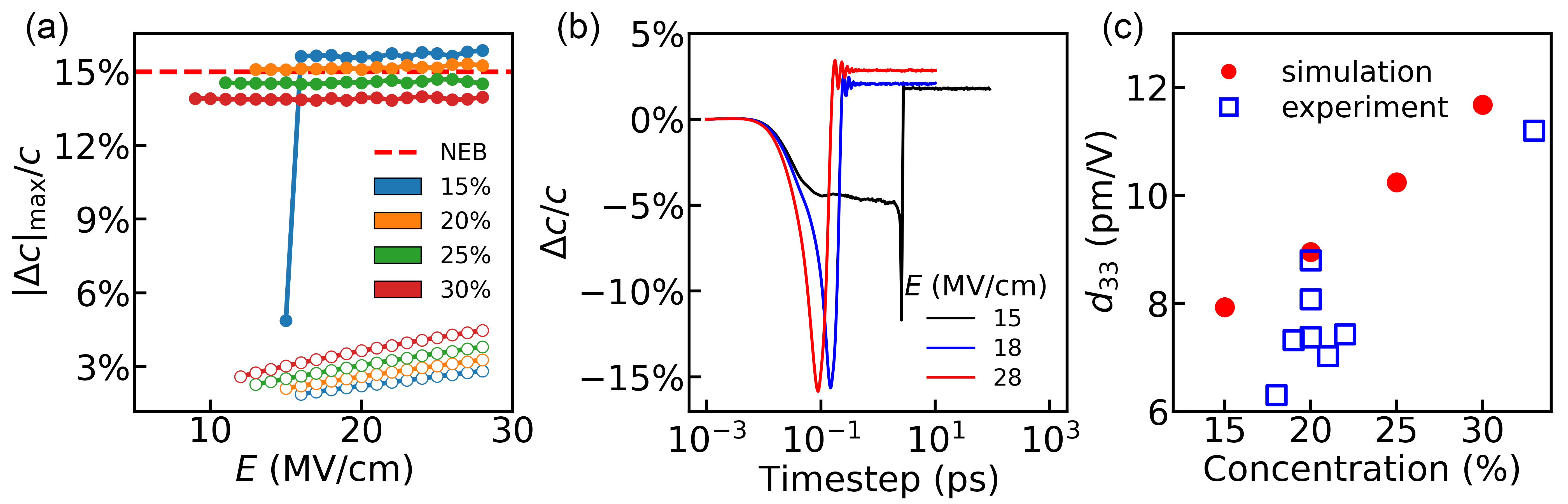}
    \caption{(a)~Maximum magnitude of the $c$-axis strain as a function of electric field $E$ for monodomain (solid symbols) and multidomain (open symbols) structures at various Sc concentrations. The red dashed line shows the SS-NEB result for Al\textsubscript{0.75}Sc\textsubscript{0.25}N in Fig.~1(c).~(b)~Time evolution of $c$-axis strain in monodomain Al\textsubscript{0.85}Sc\textsubscript{0.15}N under $E=15$ (black), $18$ (blue), and $28$ (red) MV/cm.~(c)~Comparison of simulated piezoelectric coefficient $d_{33}$ (red solid circles) with experimental values (blue open squares) \cite{patidar_2024,stefan_2019}.}
    \label{fig:strain}
\end{figure}

Figure~\ref{fig:strain}(a) summarizes the evolution of the $c$-axis lattice parameter during polarization switching for four Sc concentrations between 15\% and 30\%. The data include results from two initial structural configurations: monodomain (solid markers) and multidomain (open markers). For comparison, the NEB results for $\text{Al}_{0.75}\text{Sc}_{0.25}\text{N}$ are shown as the red dashed line. We noted that the monodomain configuration remained at a relatively high strain level across the applied electric field range. This behavior was consistent with our SS-NEB results in $\text{Al}_{0.75}\text{Sc}_{0.25}\text{N}$, except near the critical switching field $E_s$, where a pronounced drop in strain was observed. This sudden strain decrease manifested as the distinctive multi-stage behavior (Fig.~\ref{fig:strain}(b)): the system first underwent a modest contraction---much smaller than at higher fields---followed by a pronounced plateau during which this compressed state persisted. This plateau was subsequently followed by a slow, gradual elongation of the $c$-axis. This sustained, moderate strain helped explain the substantial reduction in switching‐induced $c$-axis strain observed at low fields in Fig.~\ref{fig:strain}(a), and it was unique to the 15\% Sc composition. Such a switch in strain response reflected a change in switching mechanism: rather than collective switching, the system adopted a more progressive, distortion‐assisted pathway akin to previously proposed strategies where small lattice distortions facilitate barrier reduction \cite{akiyama_2025}. The change in switching mechanism was further supported by the variation in slope obtained from the Merz’s law fitting in Fig.~\ref{fig:switch_feature}(b).

In contrast, the multidomain configurations exhibited significantly lower strain levels compared to the monodomain case. Moreover, the strain magnitude increased monotonically with the applied electric field, which was consistent with the intrinsic piezoelectric response of the material. This difference highlighted the fundamental distinctions in strain evolution mechanisms between uniform (monodomain) and non-uniform (multidomain) switching processes.

Under a given electric field, we found that the fully switched structures exhibited the same magnitude of strain along the $c$-axis. This strain, however, varied systematically with the applied field, which we attributed to the converse piezoelectric effect. To quantify this effect, we evaluated the effective $d_{33}$ coefficient by correlating the lattice strain with the electric field strength, and compared the results with experimental data (Fig.~\ref{fig:strain}(c)). Our calculations yielded slightly larger values than those reported experimentally \cite{patidar_2024,stefan_2019}. We ascribed this discrepancy to the use of the NPT ensemble, which allowed the structural stress to be fully relaxed, whereas real materials experience residual stress due to mechanical constraints.

\subsection{Characteristic Features of Polarization Switching}

\begin{figure}[htbp]
    \centering
    \includegraphics[width=0.6\textwidth]{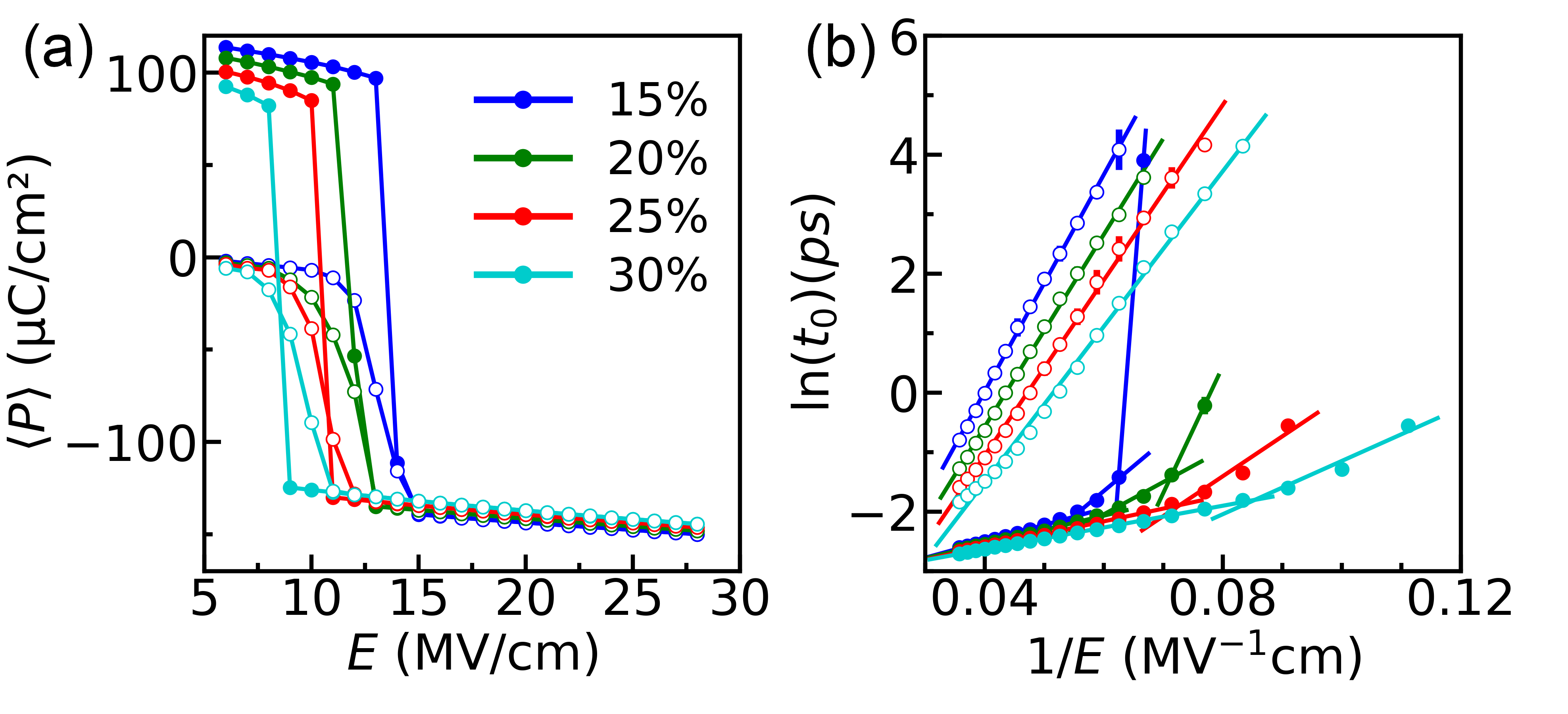}
    \caption{Switchable polarization and switching dynamics in $\mathrm{Al}_{1-x}\mathrm{Sc}_{x}\mathrm{N}$.~(a)~Final polarization values as a function of electric field for monodomain (solid symbols) and multidomain (open symbols) configurations across different Sc concentrations.~(b)~Characteristic switching time \( t_0 \) shows a linear dependence of \(\ln(t_0)\) on \(1/E\), consistent with Merz’s law. Symbols follow the same convention as in (a).}
    \label{fig:switch_feature}
\end{figure}

Figure~\ref{fig:switch_feature}(a) shows the final polarization values under various electric fields for monodomain (solid symbols) and multidomain (open symbols) structures across different Sc concentrations. A well-defined switching field $E_s$ could be identified in all cases, which systematically decreased with increasing Sc content—consistent with the results shown in Fig.~2(b) of the main text.

When switching was completed within the 1 ns simulation window, the final polarization values for monodomain and multidomain configurations converged to similar values at a given Sc concentration and field strength. This indicates that the fully switched state is mainly determined by intrinsic factors such as Sc content and applied field, with only a weak dependence on simulation cell size. Such kinetic constraints became more evident when examining the switching dynamics. As discussed in the main text, polarization reversal in multidomain structures was mediated by domain-wall propagation, which tended to be slower than nucleation-driven switching. Incomplete switching observed below $E_s$ was thus attributed to the sluggish expansion of existing domains within the limited simulation time.

To further characterize the switching process, we analyze the field dependence of the characteristic switching time $t_{0}$ for monodomain and multidomain systems in Fig.~\ref{fig:switch_feature}(b). The data roughly followed Merz’s law behavior, with $\ln(t_0)$ scaling linearly with $1/E$. However, the monodomain results exhibited multiple linear regimes instead of a single slope. We attribute this multi-segment behavior to different dominant switching mechanisms at different field strengths. At high fields, switching was abrupt and collective, characterized by large, sudden lattice strain and collective reversal. At intermediate electric fields, the applied field was insufficient to directly overcome the switching barrier associated with the collective mechanism. Instead, the system first underwent a gradual $c$-axis contraction, which likely served to lower the energy barrier by inducing lattice strain. Once the barrier was sufficiently reduced, collective switching ensued, indicating a two-step process involving pre-conditioning of the lattice. In particular, in the monodomain $\text{Al}_{0.75}\text{Sc}_{0.25}\text{N}$ structure at 15 MV/cm, collective switching could not be sustained even after the initial lattice deformation. Indeed, the switching process exhibited an initial $c$-axis contraction followed by polarization reversal via domain nucleation and growth (see Fig.~\ref{fig:strain}(b)).

\subsection{Time-Dependent Evolution of Coordination}

To gain more atomistic insights into the switching mechanism, we analyzed the time-dependent evolution of the coordination for individual cation species. In ideal wurtzite-type AlScN alloys, both Al and Sc atoms are tetrahedrally coordinated by N atoms (i.e., AlN$_4$ and ScN$_4$). Consequently, the deviation from this four-fold coordination serves as a sensitive indicator of local bond breaking and reforming events during switching. Therefore, we define the fractions of four-fold coordinated Al and Sc atoms, denoted as $f_{\text{AlN}_4}(t)$ and $f_{\text{ScN}_4}(t)$, respectively. These values are calculated by normalizing the instantaneous count of four-fold coordinated atoms by the initial number of upward-polarized Al (or Sc) atoms.

\begin{figure}[htbp]
  \centering
  \includegraphics[width=0.3\textwidth]{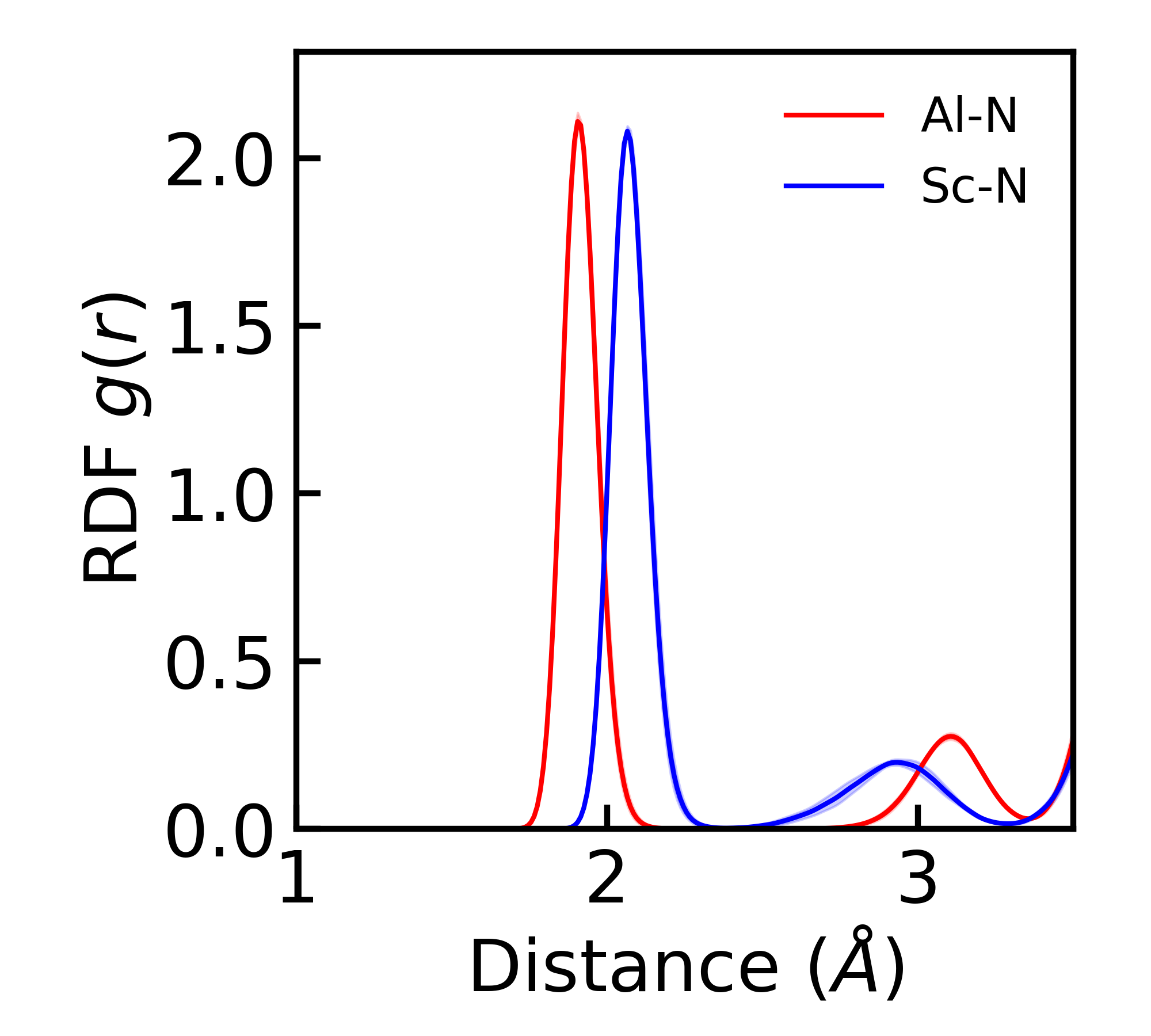}
  \caption{Radial distribution functions (RDF) of cation-nitrogen pairs in $\mathrm{Al}_{1-x}\mathrm{Sc}_{x}\mathrm{N}$, averaged across various Sc concentrations (15\%, 20\%, 25\%, and 30\%). The red and blue curves represent the Al-N and Sc-N pairs, respectively.}
  \label{fig:rdf}
\end{figure}

Specifically, the criteria for bond formation and breaking were determined based on the radial distribution functions (RDF) of Al-N and Sc-N pairs, averaged across various Sc concentrations, as shown in Fig.~\ref{fig:rdf}. Statistical analysis of the RDFs yielded a first coordination peak at $1.91$~\AA{} and a first local minimum at $2.33$~\AA{} for Al-N pairs, whereas for Sc-N pairs, these features appeared at $2.07$~\AA{} and $2.38$~\AA{}, respectively. Based on these features, the bond breaking cutoff ($r_{\text{break}}$) is defined slightly beyond the first local minimum ($r_{\text{min}} + \delta$) to fully encompass the first coordination shell. To eliminate spurious fluctuations caused by thermal noise (i.e., rapid flickering of bond states), we adopted a hysteresis scheme by setting a distinct, tighter threshold for bond formation ($r_{\text{form}}$). For the shorter and stiffer Al-N bonds, we applied a conservative buffer ($\delta \approx 0.05$~\AA{}), resulting in $r_{\text{break}}=2.38$~\AA{} and $r_{\text{form}}=2.18$~\AA{}. Conversely, for the longer and softer Sc-N bonds, a larger tolerance ($\delta \approx 0.15$~\AA{}) was necessary to accommodate their broader distribution, yielding relaxed cutoffs of $r_{\text{break}}=2.53$~\AA{} and $r_{\text{form}}=2.28$~\AA{}. 

\begin{figure}[htbp]
  \centering
  \includegraphics[width=0.9\textwidth]{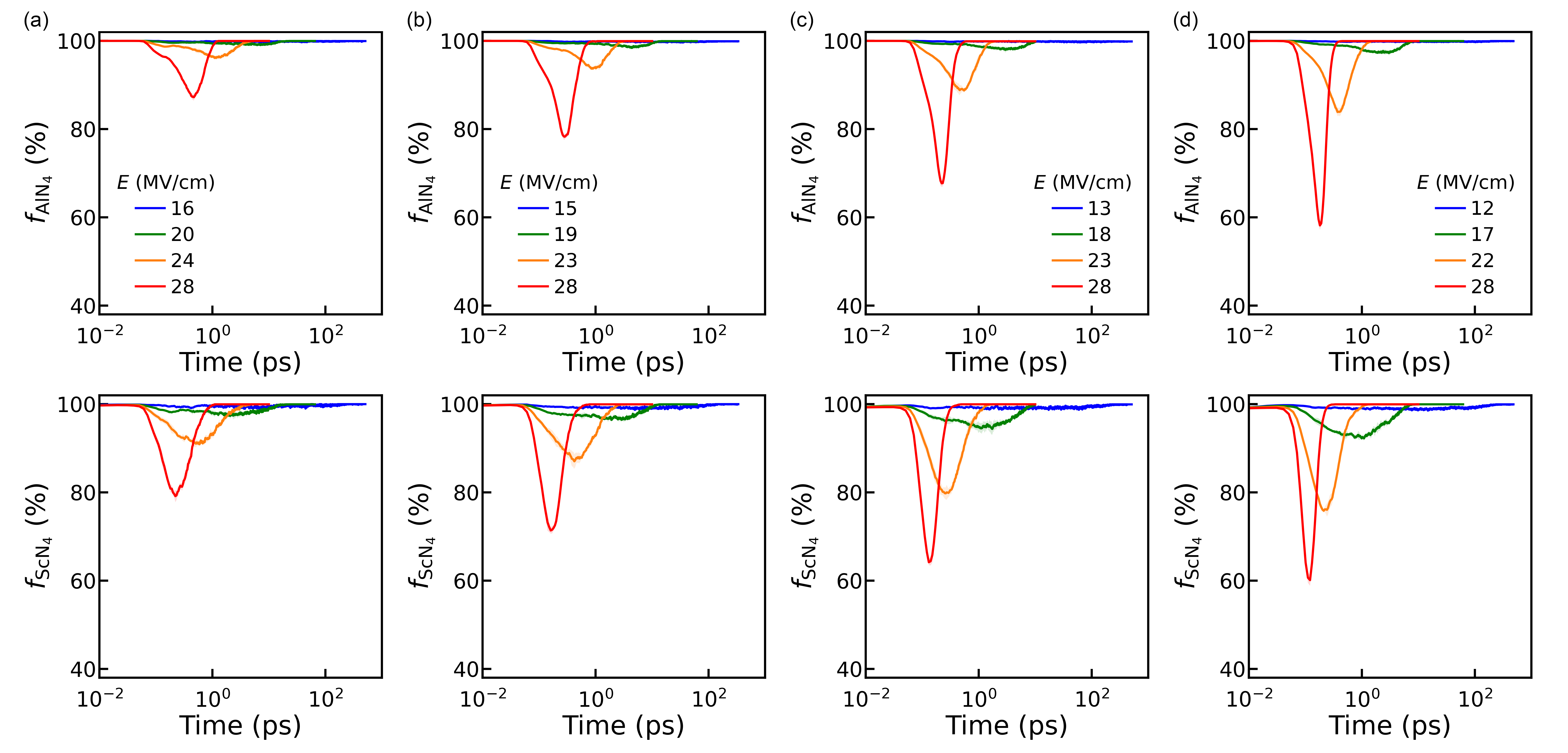}
  \caption{Time evolution of the fraction of four-fold coordinated atoms, $f_{\mathrm{AlN}_4}$ (top panels) and $f_{\mathrm{ScN}_4}$ (bottom panels), during polarization switching in $\mathrm{Al}_{1-x}\mathrm{Sc}_{x}\mathrm{N}$ under various electric fields ($E$). The columns (a)--(d) correspond to Sc concentrations of $x = 0.15$, $0.20$, $0.25$, and $0.30$, respectively. In each panel, the four curves represent different electric fields, color-coded in ascending order as blue, green, orange, and red.}
  \label{fig:cn4}
\end{figure}

Figure~\ref{fig:cn4} illustrates the time evolution of $f_{\mathrm{AlN}_4}$ and $f_{\mathrm{ScN}_4}$ across various Sc concentrations and electric fields. At low electric fields (blue curves), both coordination fractions remained stable over time. This stability implies that bond breaking and reforming were mainly localized near the domain wall, consistent with a propagation-dominated switching mechanism. Conversely, at high electric fields (red curves), a pronounced valley in both $f_{\mathrm{AlN}_4}$ and $f_{\mathrm{ScN}_4}$ was observed. This significant drop corresponded to widespread bond breaking events, indicative of a nucleation-dominated process. At intermediate fields, the switching likely involved a competition between these two mechanisms. Notably, the minimum of $f_{\mathrm{ScN}_4}$ preceded that of $f_{\mathrm{AlN}_4}$ under the same conditions. This temporal offset confirms that Sc atoms serve as preferential nucleation sites and drive the reversal of the surrounding atoms. Additionally, increasing the Sc concentration ($x$) led to a deeper and earlier minimum. These trends suggest that Sc incorporation effectively accelerates the overall switching process.

\subsection{Determination of Critical Transition Fields via Potential Energy Analysis}

\begin{figure}[htbp]
  \centering
  \includegraphics[width=0.6\textwidth]{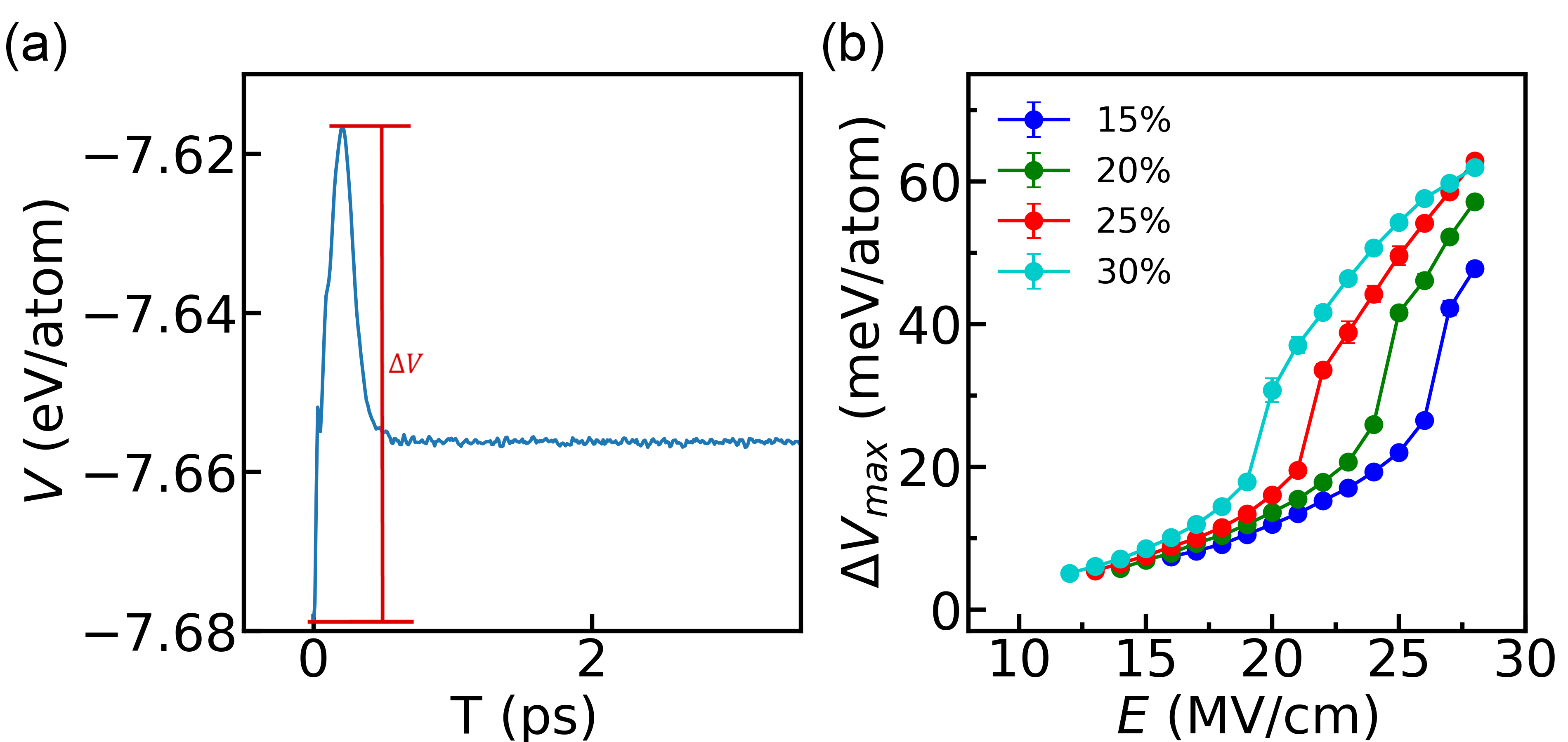}
  \caption{Electric-field dependence of switching energetics and dynamics.~(a)~Illustration of the potential energy fluctuation ($\Delta V$) determination.~(b)~$\Delta V$ as a function of the applied electric field for multidomain structures. Different Sc concentrations are distinguished by color.}
  \label{fig:energy_fluctuation}
\end{figure}

Nucleation-dominated processes involve a large number of atoms in the transition state, leading to pronounced energy fluctuations. Conversely, domain-wall propagation proceeds via localized atomic movements, resulting in comparatively minor energy changes. Based on this physical insight, we introduce the maximum potential energy difference ($\Delta V_{max}$) as an energetic descriptor. Based on the potential energy profile shown in Fig.~\ref{fig:energy_fluctuation}(a), we calculated $\Delta V$ as the difference between a local peak and its preceding valley. The maximum $\Delta V$ observed during the trajectory was labeled $\Delta V_{max}$.

\begin{table}[h]
    \centering
    \renewcommand{\thetable}{S1}
    \caption{The determined critical electric fields ($E_c$) distinguishing the KAI and SNNG mechanisms for different Sc concentrations.}
    \begin{tabular}{ccccc}
        \hline\hline
        ratio(\%) & 15 & 20 & 25 & 30 \\ 
        \hline       
        $E_c$ (MV/cm) & 26-27 & 24-25 & 21-22 & 19-20 \\ 
        \hline\hline
    \end{tabular}
\end{table}

\begin{figure}[htbp]
    \centering
    \includegraphics[width=0.3\textwidth]{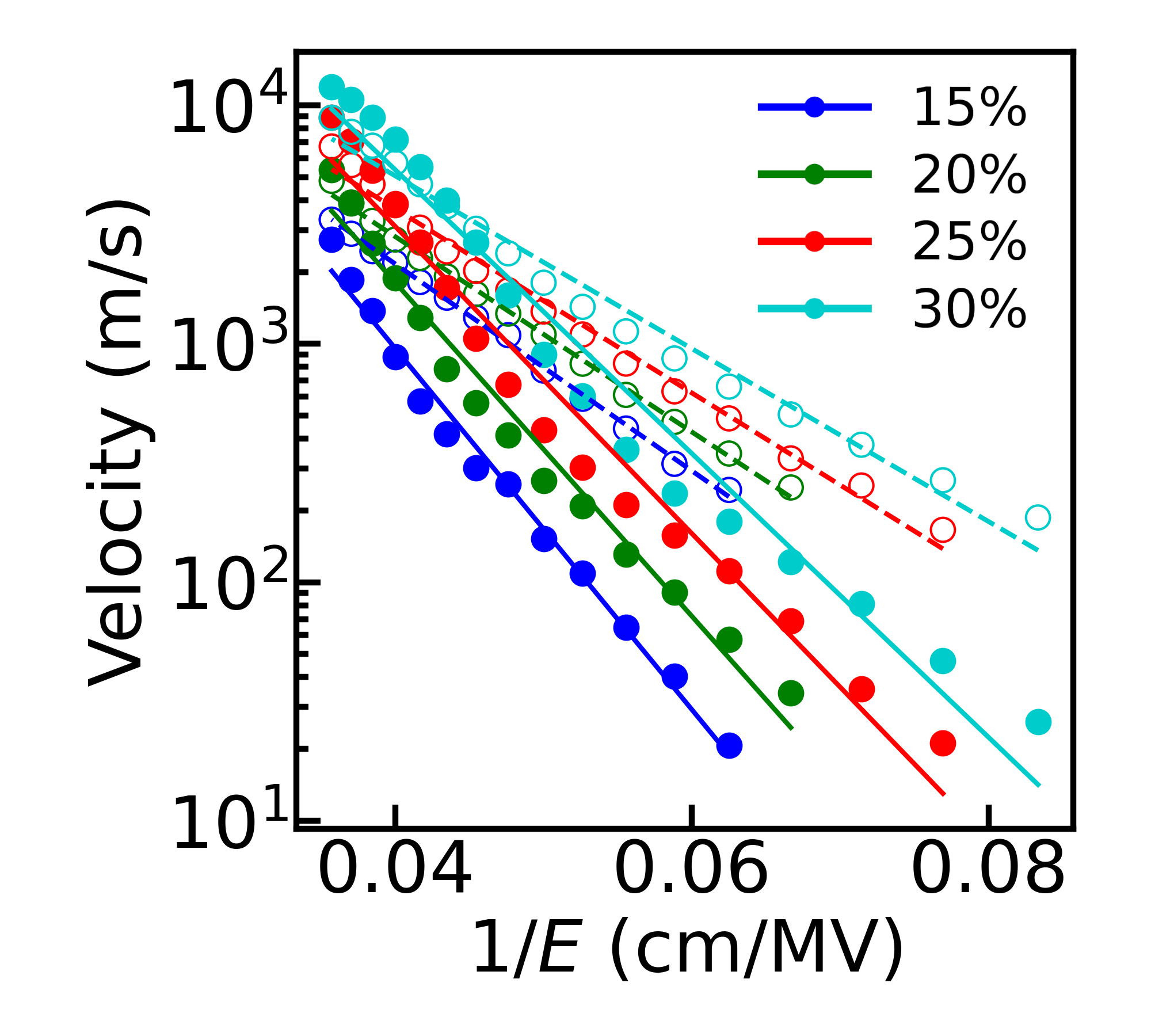}
    \caption{Domain-wall propagation velocity as a function of $1/E$ for different Sc concentrations. Solid symbols represent motion along the $x$ direction, and open symbols denote motion along the $z$ direction. Colors correspond to different Sc compositions.}
    \label{fig:dw_vel}
\end{figure}

As shown in Fig.~\ref{fig:energy_fluctuation}(b), we performed an analysis of multidomain structures to validate this descriptor. The multidomain data clearly exhibited two distinct linear regimes, marking the transition between two mechanisms. In the low-field linear regime, $\Delta V_{max}$ followed a gentle linear growth trend. This corresponded to the KAI model, where switching is dominated by domain-wall propagation. This linear dependence was consistent with domain wall kinetics, where the propagation velocity scales linearly with the electric field (see Fig.~\ref{fig:dw_vel}). As the field increased beyond a certain threshold, $\Delta V_{max}$ transitioned into a second linear regime characterized by both an abrupt elevation in magnitude and a significantly steeper slope. This regime corresponded to the SNNG model, where the rapid increase in energy fluctuations is driven by high-density nucleation events. By identifying the crossover point between these two linear regimes, we determined the critical electric fields for varying Sc concentrations (Table.~S1). 

\section{Model Fitting of Polarization Dynamics}

For three representative cases of each switching mechanism observed in the multidomain structures, the calculated time dependence of the domain ratio ($d_{dw}$) was fitted using three widely adopted kinetic models: the Kolmogorov-Avrami-Ishibashi (KAI) model \cite{avrami_1940,ishibashi_1971,kolmogorov_1937}, the nucleation-limited-switching (NLS) model \cite{tagantsev_2002} and the simultaneous non-linear nucleation and growth (SNNG) model \cite{yazawa_2023}.

Traditionally, model fitting is performed using the normalized polarization change, $\Delta P / (2 P_s)$. However, since the material exhibits a significant piezoelectric response, polarization changes include both true switching and elastic lattice contributions. To eliminate the influence of the piezoelectric effect, we instead used $d_{dw}$ as the fitting quantity, with $f = d_{dw} \times 2$ as the normalization factor. This approach isolated the genuine switching dynamics from lattice-mediated polarization variations, allowing a more accurate assessment of the kinetic behavior.

Both KAI and NLS models can be expressed in a general form as~\cite{tagantsev_2002}
\begin{equation}
f = \int_{-\infty}^{+\infty} \left[ 1 - \exp \left\{ -(t/t_0)^n \right\} \right] F(\log t_0) d(\log t_0).
\end{equation}
The difference between these two models lies in the form of the distribution function $F(\log t_0)$. In the NLS model, $F(\log t_0)$ follows a Lorentzian distribution given by \cite{jo_2007}:
\begin{equation}
F(\log t_0) = \frac{A}{\pi} \left[ \frac{w}{(\log t_0 - \log t_1)^2 + w^2} \right],
\end{equation}
where $A$ is a normalization constant, $t_1$ is the center of the distribution and $2w$ is the full width at half maximum (FWHM). For the KAI model, the $F(\log t_0)$ distribution function reduces to a Dirac delta function $\delta(\log t_0 - \log t_k)$, where $t_k$ represents the characteristic switching time \cite{jo_2007, lu_2024, wei_2021}. Substituting this function into Eq.~(S2) yields the KAI expression \cite{avrami_1940,ishibashi_1971,kolmogorov_1937}:
\begin{equation}
f = \left[1 - \exp\left\{ -\left( {t}/{t_k} \right)^n \right\} \right]
\end{equation}
where $n$ is the Avrami exponent, reflecting the dimensionality of domain growth and nucleation rate. The SNNG model is described by \cite{yazawa_2023}:
\begin{equation}
f = \left[ 1 - \exp\left\{ -2\pi d v^2 N(\infty) \alpha m \int_0^t \tau^{m-1} (t - \tau)^2 \exp(-\alpha \tau^m) \, d\tau \right\} \right]
\end{equation}
where $d$ is the film thickness, $v$ is the growth speed, $N(\infty)$ is the saturated density of nuclei, $\alpha$ is the constant nucleation rate and $m$ is the nucleation rate peak shape parameter. The fitting results are shown in Figs.~\ref{fig:model_fit}-\ref{fig:model_n_fwhm}.

\clearpage

\begin{figure}[htbp]
    \centering
    \includegraphics[width=\textwidth]{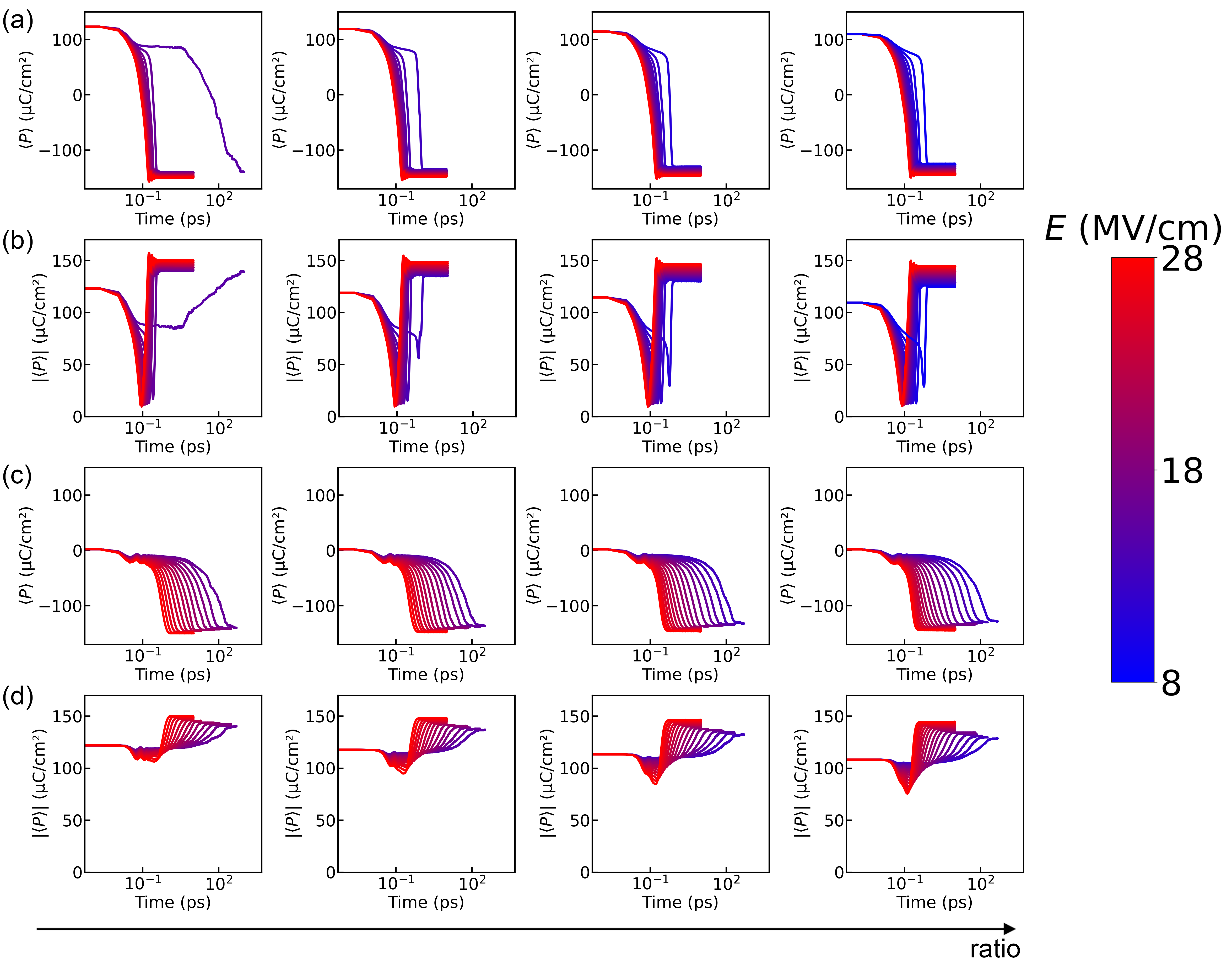}
    \caption{Time evolution of polarization under various electric fields for both monodomain and multidomain configurations at different Sc concentrations. The 4$\times$4 panel array is organized such that each column corresponds to a different Sc concentration (15\%, 20\%, 25\%, and 30\%), and each row represents a different dataset category:~(a)~total polarization of monodomain structures;~(b)~local polarization (averaged over spatial regions) in monodomain structures;~(c)~total polarization of multidomain structures; and (d)~local polarization in multidomain structures. Within each subplot, multiple colored curves represent independent simulations under different external electric fields, and only those simulations that result in complete polarization reversal are included. The color bar on the right encodes the magnitude of the applied electric field, increasing from bottom to top.
    }
    \label{fig:pol_all}
\end{figure}

\begin{figure}[htbp]
    \centering
    \includegraphics[width=0.5\textwidth]{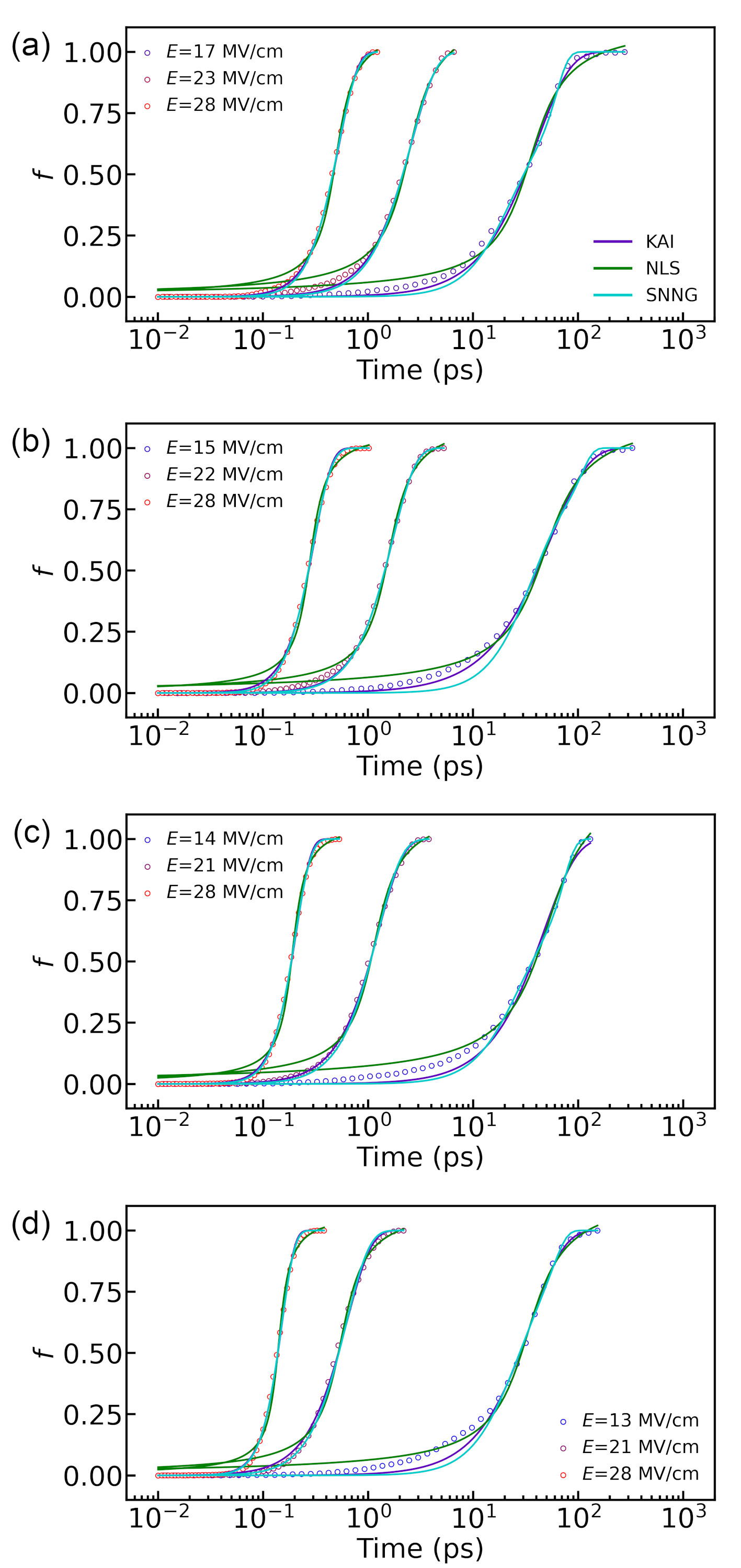}
    \caption{Fitting of polarization dynamics with different models.~(a–d)~Polarization versus time at 15\%, 20\%, 25\%, and 30\% Sc concentrations, respectively. 
    Solid lines represent model fits (KAI, NLS, SNNG), and dots show simulation data under varying electric fields.
    }
    \label{fig:model_fit}
\end{figure}

\begin{figure}[htbp]
    \centering
    \includegraphics[width=\textwidth]{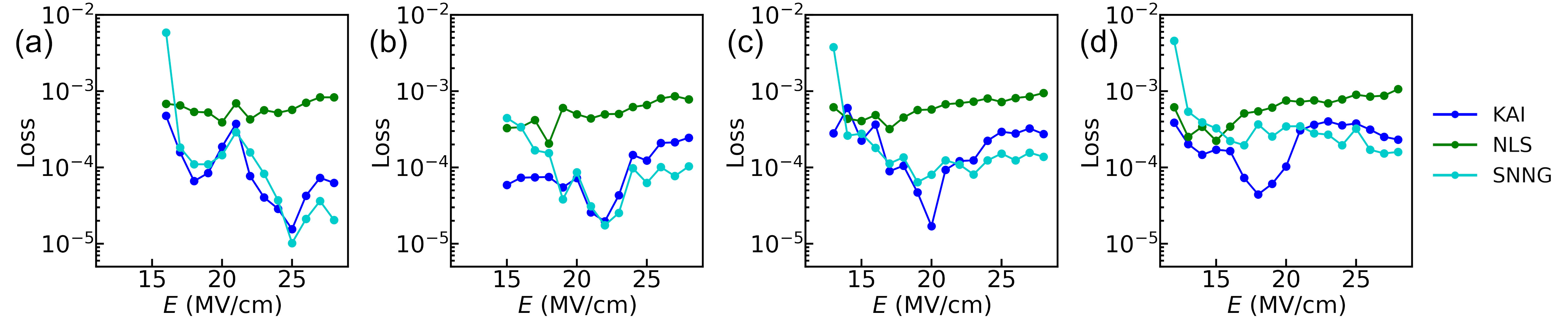}
    \caption{Loss values for different models.~(a–d)~Loss function values for KAI, NLS and SNNG models at 15\%, 20\%, 25\%, and 30\% Sc concentrations, respectively. Each marker represents one electric field condition.}
    \label{fig:model_loss}

    \vspace{1cm}

    \includegraphics[width=0.6\textwidth]{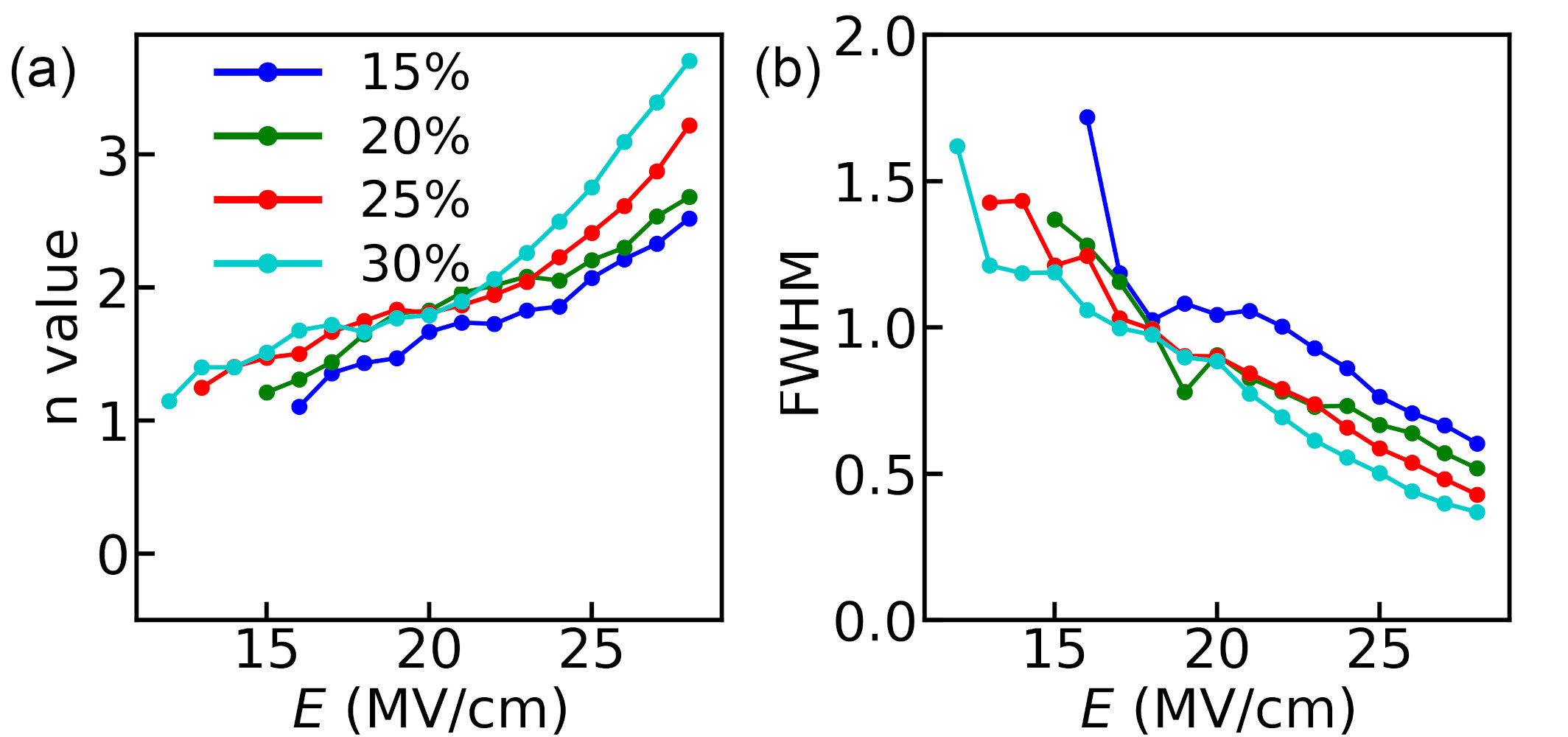}
    \caption{Comparison of KAI and NLS models across electric fields.~(a)~Avrami exponent $n$ from the KAI model as functions of electric field.~(b)~FWHM of the log-time switching profile from the NLS model. Different curves correspond to different Sc concentrations (15\%, 20\%, 25\%, 30\%).}
    \label{fig:model_n_fwhm}
\end{figure}

\clearpage

\putbib[reference]
\end{bibunit}

\end{document}